\documentclass[]{aa}
\pdfoutput=1
\usepackage{natbib}
\usepackage{graphicx}
\usepackage{color}
\usepackage{amsmath}
\usepackage{url}
\usepackage{txfonts}
\usepackage{enumitem}
\usepackage{bm}

\newcommand\bb[1] {   \mbox{\boldmath{$#1$}}  }
\newcommand\del{\bb{\nabla}}
\newcommand\bcdot{\bb{\cdot}}

\newcommand{\ba}{\begin{eqnarray}}
\newcommand{\ea}{\end{eqnarray}}
\newcommand{\be}{\begin{equation}}
\newcommand{\ee}{\end{equation}}

\begin{document}

\title{ Angular
  momentum transport and large eddy simulations in magnetorotational turbulence: the small Pm limit.}
\author{Heloise Meheut \inst{1}, S\'ebastien Fromang \inst{1}, 
  Geoffroy Lesur \inst{2,3}, Marc Joos  \inst{1} and Pierre-Yves Longaretti \inst{2,3}} 

\institute{Laboratoire AIM, CEA/DSM-CNRS-Universit\'e Paris 7, Irfu/Service d'Astrophysique, CEA-Saclay, 91191 Gif-sur-Yvette, France\and
Univ. Grenoble Alpes, IPAG, F-38000 Grenoble, France \and
CNRS, IPAG, F-38000 Grenoble, France\\ \email{heloise.meheut@cea.fr}} 

\date{Accepted ; Received ; in original form ;}

\abstract
{Angular momentum transport in accretion discs is often believed to be due to
  magnetohydrodynamic turbulence mediated by the magnetorotational
  instability (MRI). Despite an abundant literature on the MRI,
  the parameters governing the saturation amplitude of the turbulence
  are poorly understood and the existence of an asymptotic
  behavior in the Ohmic diffusion regime
  is not clearly established.}
{We investigate the properties of the turbulent state in the
  small magnetic Prandtl number limit. Since this is
  extremely computationally expensive, we also study the relevance and range of
  applicability of the most common subgrid scale models for this problem.
}
{
 Unstratified shearing boxes simulations are
  performed both in the compressible and
  incompressible limits, with a resolution up to 800 cells per disc
  scale height. The latter constitutes the largest resolution ever
  attained for a simulation of MRI turbulence. Different magnetic field
  geometry and a large range of dimensionless dissipative
  coefficients are considered. 
  We also systematically investigate
  the interest of using large eddy simulations (LES).
  }
{In the presence of a mean magnetic field threading the domain,
  angular momentum transport converges to a finite value in the small
  $Pm$ limit. When the mean vertical field amplitude is such that
  $\beta$, the ratio between the thermal and magnetic pressure, equals
  $10^3$, we find $\alpha \sim 3.2 \times 10^{-2}$ when $Pm$ approaches
  zero. In the case of a mean toroidal field for which $\beta=100$, we
  find $\alpha \sim 1.8 \times 10^{-2}$ in the same limit. Both implicit
  LES and Chollet-Lesieur closure model reproduces these results
  for the $\alpha$ parameter and the power spectra. A reduction in 
  computational cost of a factor at least $16$ (and up to $256$) is achieved
   when using such methods.
  }
{MRI turbulence operates efficiently in the small Pm
  limit provided there is a mean magnetic field. Implicit LES offers a practical and efficient mean of
  investigation of this regime but should be used with care,
  particularly in the case of a vertical field. Chollet-Lesieur 
  closure model is perfectly suited for simulations done with a spectral
  code.}

\authorrunning{Meheut et al.}
\titlerunning{MRI turbulence in the small Pm limit}
\maketitle

%%%%%%%%%%%%%%%%%%%%%%%%%%%%%
%%%%%%%%%%Introduction
\section{Introduction}
\label{sec:intro}

Determining the rate of angular momentum transport in accretion discs
is considered as one of the key unsolved astrophysical questions.  
Accretion is believed to be at the origin of the radiation emitted by
some of the most luminous sources in the universe from active galactic
nuclei to X-ray binaries and is also a major process at work
during planet formation in protoplanetary discs
\citep{FRA02}. In addition, accretion discs 
are ubiquitous in the universe and affect the dynamics, evolution and
appearance of multiple astrophysical objects at all spatial and energy
scales. The accretion rate can be indirectly constrained by the
luminosity of high-energy sources or the life-time of protoplanetary
discs, and, if the transport process is modeled by an $\alpha$ viscosity
\citep{SHA73}, such an estimate gives $0.01<\alpha<0.4$ depending on
the system considered \citep{KPL07}. 

The physical origin of angular momentum transport has to be understood
to explain such an efficient radial angular momentum
transport. Currently, the most widely accepted mechanism is magnetohydrodynamic (MHD) 
turbulence induced by the non-linear evolution of the
magneto-rotational instability \citep[MRI,][]{BH91}. That instability
along with its nonlinear development has been extensively studied
in the last two decades \citep{BH98,B03,F13}. Using
local simulations performed in the framework of the shearing box, 
\citet{HGB95} quickly established that the MRI develops into
vigorous MHD turbulence that efficiently transports angular momentum
outward, a result that was later confirmed to be independent of the
field geometry \citep{HGB96} or to the background disc
stratification \citep{BNS95,SHG96}. Only
recently the sensitivity of MRI--driven MHD turbulence saturation
to small scale dissipation in such idealized simulations has been
identified: when magnetic field diffusion is dominated by an ohmic
resistivity $\eta$, $\alpha$ is an increasing function of the magnetic
Prandtl number $Pm$, the ratio between the kinematic viscosity $\nu$
and $\eta$. This is the case both in the presence of a mean vertical
magnetic field \citep{LL07} and of a mean azimuthal magnetic field
\citep{SH09}. Such a ``Pm--effect'' has
a significant impact on the rate of angular momentum transport
measured in homogeneous shearing boxes simulations: in the
case of a mean vertical field such that the plasma $\beta$
parameter (defined in section \ref{sec:model_params}) amounts to $10^3$, \citet{LL10} showed that 
$\alpha$ varies by a factor of about $5$ when $Pm$ only varies between
$1/4$ and $4$, without any sign of the relation flattening at either
range.
The dynamo case (i.e. no mean magnetic field
threading the computational domain) also displays a large sensitivity
to $Pm$ \citep{FPL07,SH09,GGS09,BCF11} that
persists when density stratification is included \citep{SHB11}. In
this dynamo regime and without stratification, the effect is even stronger and dynamo action is suppressed
for $Pm$ values smaller than unity, in which case the flow remains
laminar, i.e. $\alpha$ goes to zero. A clear understanding of the
origin of the effect of $Pm$ on the rate of angular momentum transport
is still lacking and is currently a matter of active research. The
first results, based on methods borrowed from the fluid community
\citep{HRC11,RRC13} are promising \citep{RRC15}
and should be extended to more realistic geometries and dimensionless
numbers.

Taken together, the results described above question the relevance of
MRI--driven MHD turbulence as the dominant transport mechanism in
accretion discs where the Prandtl number can be orders of magnitude
lower ($Pm \ll 1$) than has been explored in published simulations
\citep{BH08}. There is clearly the possibility that $\alpha$ becomes
vanishingly small as $Pm$ decreases to small, but astrophysically
relevant values. The first goal of this paper is to investigate that
asymptotic behavior by means of high resolution simulations performed
in the homogeneous shearing box. In doing so, we will leave aside the
dynamo case and focus on vertical and azimuthal mean field
configurations. Using high resolution simulations (such that the
coverage in $Pm$ now extends over more than two orders of magnitudes, from
$10^{-2}$ to $4$), we will show that $\alpha$ asymptotically
converges to a well defined finite value in both cases. However, 
 the computational cost associated with such simulations is
extremely large (for example, $15$ millions CPU hours are needed to
complete the $1000^3$ simulation we describe in
section~\ref{sec:GChallenge} on a BlueGen/Q machine ranked $42$nd
 on the Top500 supercomputing website in
november 2014 \footnote{see
http://www.top500.org/list/2014/11/}). In practice, such a large cost is
prohibitive if one wants to perform a parameter survey in the
asymptotic regime with additional physics included and 
simply prevents global simulations to be performed in that limit. The
second goal of this paper is thus to investigate the possibility to
use sub--grid scale models as a mean to reduce the computational cost
associated with MHD turbulence simulations in the small $Pm$ limit.

Recent years have seen significant progress in our
understanding of the consequences of ambipolar diffusion
\citep{BS11,BS13,SBA13} and the Hall effect
\citep{KL13,LKF14,B15}, both of which are particularly
important in shaping the structure of protoplanetary discs. In this
paper we will restrict our attention to Ohmic resistivity as the 
sole source of magnetic field dissipation. Such a regime is
relevant to describe the very inner parts of protoplanetary discs (at
stellar distances of a few tens of an AU) but also cataclysmic variables (CV) discs and the
outer parts of X-ray binaries discs \citep{BH08}. In principle, the
analysis we present here should also be carried for such cases where
ambipolar diffusion or the Hall effect are the dominant magnetic field
diffusion processes. 

The plan of the paper is as follows. In the following section,
the equations and numerical methods are given. Section
\ref{sec:results} presents our results, focusing first on simulations
that explicitly resolve the small dissipative scales
(section~\ref{sec:DNS}) and then on two different methods to
perform Large Eddy Simulations (LES) in section~\ref{sec:LES}. We
finally conclude and discuss the implications of our work in
section~\ref{sec:conclusion}. 

%%%%%%%%%%%%%%%%%%%%%%%%%%%%%
%%%%%%%%%%Methods
\section{Methods}

%###############Equations
\subsection{Equations and notations}

In this paper, we use the shearing box approximation
\citep{HGB95}. Namely, we 
compute the evolution of the fluid in a small box centered at a radius
$r_0$ of the disc and rotating at the same velocity as the fluid at
$r=r_0$. As the size of the box is small compared to the radial
position, the curvature terms of the standard fluid equations can be
simplified \citep{HB92}. We thus use Cartesian  coordinates $(x,y,z)$
with units vectors ($\vec i,\vec j,\vec k$). In this coordinate
system, we note $(L_x,L_y,L_z)$ the size of the computational
box. Neglecting the vertical component of the gravitational
acceleration, we solve the following set of equations: 
\ba
\partial_t\rho+\vec\nabla\cdot(\rho\vec v)&=&0 \label{eq:continuity} \\
\partial_t(\rho\vec v)+\vec\nabla\cdot(\rho\vec v\vec v-\vec B\vec
B)+\vec\nabla P_{tot}&=&2q\rho\Omega_0^2 x \vec
i-2\rho\vec\Omega_0\times\vec v\nonumber \\ +\vec\nabla\cdot\vec
T \label{eq:momentum} \\
\partial_t\vec B&=&\vec\nabla\times(\vec v\times\vec
B-\eta\vec\nabla\times\vec B) \label{eq:induction}
\ea
where $\rho$ is the fluid density, $\vec v$ its velocity in the
rotating frame, $\vec B$ is the magnetic field and $\Omega_0$ is the angular
velocity of the fluid at $r_0$, corresponding to the angular velocity
of the box. $q$ stands for the background keplerian shear and is taken
equal to $1.5$ throughout this paper. $P_{tot}$ is the total pressure,
the sum of the thermal pressure $P$ and the magnetic pressure
$B^2/2$.
Explicit
dissipation is accounted for through the Ohmic resistivity $\eta$ and
the kinematic viscosity $\nu$ that enters in the viscous stress tensor
$\bb{T}$ defined as 
\be
T_{ij}=\rho\nu
\left(\partial_{j}v_i+\partial_{i}v_j-\frac{2}{3}\delta_{ij}\vec\nabla\cdot\vec
v \right) \, .
\ee
The amplitude of viscosity and resistivity are set using the
magnitude of the Reynolds number $Re$ and magnetic Reynolds number
$Rm$ with the following relations
\be
Re = \frac{\Omega_0 L_z^2}{\nu} \, ,
Rm = \frac{\Omega_0 L_z^2}{\eta}
\nonumber
\ee
which can also serve as an alternative definition for the magnetic
Prandtl number $Pm$ already given in the introduction: 
\be
Pm=\frac{\nu}{\eta}=\frac{Rm}{Re} \, .
\ee

We performed simulations that consider two different flavors of
the system of equations (\ref{eq:continuity}), (\ref{eq:momentum}) and
(\ref{eq:induction}). First, we solve the above equations in the
incompressible limit. In this case, the density is constant and equation (\ref{eq:continuity}) reduces to 
$\del \bcdot \bb{v}=0$. We used the code SNOOPY in that case (see
section~\ref{sec:snoopy}). In the second type of simulations, we solve
the full set of equations (and refer to that case as the compressible
simulations) using the code RAMSES (section~\ref{sec:ramses}). We
briefly describe below the two codes and the specificities of each
set of simulations.

%###############SNOOPY
\subsection{Incompressible simulations: the SNOOPY code}
\label{sec:snoopy}

The SNOOPY code is a pseudo spectral code that solves the incompressible
MHD equations in a Fourier basis. It uses a low-storage 3${}^\mathrm{rd}$ order
Runge-Kutta integrator and works in a sheared frame comoving with the
mean flow, which is equivalent to a 3${}^\mathrm{rd}$ order Fargo scheme \citep{M00}.
SNOOPY uses a $3/2$ antialiasing rule to eliminate the excitation of spurious modes
during the computation of quadratic nonlinearities.

The use of a sheared Fourier basis makes the boundary conditions periodic in the $y$
and $z$ directions and shear-periodic in the $x$ direction. SNOOPY conserves linear 
momentum and magnetic flux down to machine precision. Since spectral methods
are inherently diffusion-free and energy conserving, the addition of dissipation is required to mimic the damping
due to small scale dissipation processes\footnote{Note that the \emph{numerical stability} of the scheme does not require any dissipation. However,
the absence of dissipation in a turbulent flow naturally leads to thermalisation, a situation which does not occur in natural systems which always exhibit dissipation processes. Numerical dissipation is therefore required \emph{on physical grounds} to break the thermodynamic equilibrium and create the well known energy cascade picture.}. In SNOOPY, one can choose second order diffusion operators, hyperdiffusion operators or \cite{CL81} subgrid models (see section \ref{sec:CL}).

%###############RAMSES
\subsection{Compressible simulations: the RAMSES code}
\label{sec:ramses}

The RAMSES code is a finite volume code that solves the 
compressible MHD equations on a Cartesian grid \citep{T02,FHT06} using
the constrained transport algorithm \citep{EH88}. We use a version of the code
for which the grid is uniform (i.e. without the adaptive mesh
refinement). The source terms associated with the tidal potential are
included as described by \citet{SG10}. We use the HLLD
Riemann solver \citep{MK05} with monotonized central slope limiter, shearing box
boundary conditions in the $x$ direction \citep{HGB95} and
periodic boundary conditions in the $y$ and $z$ directions. For
extended box sizes such as used in the mean vertical field case (see
section~\ref{sec:model_params}), it has been found that radially
variable numerical dissipation causes the turbulent stress to vary
accross the box. To avoid that problem, we used the FARGO 
algorithm \citep{M00,SG10} in that case, which also
improves the efficiency of the code through an increase in the
timestep.  

Throughout this paper, we use an isothermal equation of state to close
the system of MHD equations in compressible simulations, such that
$P=\rho c_0^2$, where $c_0$ 
stands for the constant sound speed. We start the simulation with an
initial uniform density $\rho=\rho_0$ and choose $L_z=H=c_0/\Omega_0$
where $H$ is the disc scale height. 

%###############Models
\subsection{Models parameters}
\label{sec:model_params}

We start the simulations with a uniform initial magnetic field:
\be
\bb{B}=B_{0y}\bb{j} + B_{0z} \bb{k} \, .
\label{eq:init_b}
\ee
Two different initial magnetic configurations are considered in the
following, namely pure azimuthal field ($B_{0z}=0$) simulations and
pure vertical field simulations ($B_{0y}=0$). The strength of the magnetic field is defined using the
plasma parameter $\beta_i$ that is given by
\ba
\beta_i \equiv \left\{ 
\begin{array}{l l}
   \frac{q^2\rho_0 c_0^2}{B_{0i}^2} \, & \quad \text{in compressible runs,}\\
   \frac{(q\Omega L_z)^2}{B_{0i}^2} \, & \quad \text{in incompressible runs.}\\ \end{array} \right. \label{eq:betaiy}\\
\ea
For the mean azimuthal (resp. vertical) magnetic field simulations, we
used $\beta_y=112.5$ (resp. $\beta_z=10^3$). At the start of each
runs, small amplitude random velocity perturbations are added on the
three velocity components with an amplitude equal to $1 \%$ of the
sound speed in compressible runs. In incompressible simulations
 both velocity and magnetic random perturbations are added with an 
 amplitude of $0.1\Omega L_z$.

The size of the computational box is either $(L_z,4L_z,L_z)$ for the
simulations with a mean azimuthal magnetic field, or $(4L_z,4L_z,L_z)$
for the simulations with a mean vertical magnetic field. In the latter
case, it is indeed well known that boxes with $L_x=L_z$ artificially
enhance the importance of recurrent bursts in the flow structure
\citep{BMC08,JYK09}. In such box sizes and with such vertical
magnetic field, \citet{BS14} recently reported zonal flows.
We also found such structures in our simulations.

The resolution varies from $32$ cells 
per unit length up to $800$ cells per unit length. Such large
resolutions allow us to reach the largest Reynolds number (run
Y-C-Re85000, $Re=85000$) and the smallest Prandtl number (run
Z-I-Re40000, $Pm=0.01$) ever published. 
As larger structures are expected in the $y$-direction, we typically
use a resolution that is twice as coarse in 
that direction. 

%%%%%%%%%%%%%%%%%%%%%%%%%%%%%
%%%%%%%%%% Results
\section{Results}
\label{sec:results}

\begin{table*}[t]\begin{center}\begin{tabular}{@{}cccccccccc}\hline\hline
Model & Resolution & T$_{Run}$ & Re & Rm & Pm & $\alpha$ &
$\alpha_{Max}$ & $\alpha_{Rey}$ & $R$ \\
\hline\hline
Y-C-Re650  & $(64,128,64)$ & 100 & 650 & 2600 & 4 & $3.0 \times 10^{-2}$
& $2.4 \times 10^{-2}$ & $5.9 \times 10^{-3}$ & $4.2$ \\
Y-C-Re2600  & $(64,128,64)$ & 100 & 2600 & 2600 & 1 & $2.5 \times 10^{-2}$
& $1.9 \times 10^{-2}$ & $5.7 \times 10^{-3}$ & $3.4$ \\
Y-C-Re13000 & $(128,256,128)$ & 100 & 13000 & 2600 & 0.2 & $1.8 \times 10^{-2}$
& $1.4 \times 10^{-2}$ & $4.6 \times 10^{-3}$ & $3.0$ \\
Y-C-Re26000 & $(256,512,256)$ & 100 & 26000 & 2600 & 0.1 & $2.0 \times 10^{-2}$
& $1.5 \times 10^{-2}$ & $5.0 \times 10^{-3}$ & $3.0$ \\
Y-C-Re85000 & $(800,1600,832)$ & 35 & 85000 & 2600 & 0.03 & $1.8
\times 10^{-2}$ & $1.3 \times 10^{-2}$ & $4.6 \times 10^{-3}$ & $2.9$ \\ 
\hline
Y-ILES-C-64 & $(64,128,64)$ & 100 & - & 2600 & - & $1.9 \times 10^{-2}$ &
$1.4 \times 10^{-2}$ & $4.7 \times 10^{-2}$ & $3.0$ \\
Y-ILES-C-128 & $(128,256,128)$ & 100 & - & 2600 & - & $1.9 \times 10^{-2}$
& $1.4 \times 10^{-2}$ & $4.8 \times 10^{-3}$ & $3.0$ \\
Y-ILES-C-256 & $(256,512,256)$ & 100 & - & 2600 & - & $1.9 \times
10^{-2}$ & $1.4 \times 10^{-2}$ & $5.0 \times 10^{-3}$ & $2.8$ \\ 
\hline\hline
\end{tabular}
\caption{RAMSES runs with a mean azimuthal field. The box size is 
  $(L_x,L_y,L_z)=(L_z,4L_z,L_z)$. $T_{run}$ is given in local orbits time units.
  The table gives for each run its resolution, duration, Reynolds
  number, magnetic Reynolds number, magnetic Prandtl number, total
  stress, Maxwell stress, Reynolds stress and the ratio of the two.} 
\label{tab:runBy}
\end{center}
\end{table*}

\begin{table*}[t]\begin{center}\begin{tabular}{@{}cccccccccc}\hline\hline
Model & Resolution & T$_{Run}$ & Re & Rm & Pm & $\alpha$ &
$\alpha_{Max}$ & $\alpha_{Rey}$ & $R$ \\
\hline\hline
Z-C-Re400 & $(128,64,32)$ & 100 & 400 & 400 & 1 & $6.5 \times
10^{-2}$ & $4.0 \times 10^{-2}$ & $2.4 \times 10^{-2}$ & $1.6$ \\                     %
Z-C-Re800 & $(256,128,64)$ & 100 & 800 & 400 & 0.5 & $5.1 \times
10^{-2}$ & $2.9 \times 10^{-2}$ & $2.2 \times 10^{-2}$ & $1.3$ \\                     %test2
Z-C-Re3000 & $(512,256,128)$ & 100 & 3000 & 400 & 0.13 & $3.7 \times
10^{-2}$ & $2.0 \times 10^{-2}$ & $1.8 \times 10^{-2}$ & $1.1$ \\                     %Pm0p13
Z-C-Re8000 & $(1024,512,256)$ & 100 & 8000 & 400 & 0.05 &  $3.1 \times
10^{-2}$ & $1.5 \times 10^{-2}$ & $1.6 \times 10^{-2}$ & $1.0$ \\                     %Pm0p05
\hline
Z-ILES-C-32 & $(128,64,32)$ & 100 & - & 400 & - & $4.1 \times 10^{-2}$ &
$2.2 \times 10^{-2}$ & $1.9 \times 10^{-2}$ & $1.2$ \\                                     %
Z-ILES-C-64 & $(256,128,64)$ & 100 & - & 400 & - & $3.6 \times 10^{-2}$ &
$1.8 \times 10^{-2}$ & $1.8 \times 10^{-2}$ & $1.0$ \\                                     %test4
Z-ILES-C-128 & $(512,256,128)$ & 100 & - & 400 & - & $3.3 \times 10^{-2}$
& $1.6 \times 10^{-2}$ & $1.7 \times 10^{-2}$ & $1.0$ \\                                  %Pm0
Z-ILES-C-256* & $(1024,512,256)$ & 40 & - & 400 & - & $3.3 \times 10^{-2}$
& $1.6 \times 10^{-2}$ & $1.8 \times 10^{-2}$ & $0.9$ \\                                  %
\hline\hline
\end{tabular}
\caption{Same than table \ref{tab:runBy} but for RAMSES runs with a mean vertical field. The
   box size is $(L_x,L_y,L_z)=(4L_z,4L_z,L_z)$. *To save computational time, this model was performed by restarting model Z-C-Re8000 at $t=50$ removing explicit viscosity}
\label{tab:runBz}
\end{center}
\end{table*}

\begin{table*}[t]\begin{center}\begin{tabular}{@{}cccccccccc}\hline\hline
Model & Resolution & T$_{Run}$ & Re & Rm & Pm & $\alpha$ &
$\alpha_{Max}$ & $\alpha_{Rey}$ & $R$ \\
\hline\hline
Z-I-Re1300 & $(256,256,64)$ & $53$ & 1333 & 400 & 0.3 & $3.8 \times 10^{-2}$ & $2.8 \times 10^{-2}$ & $9.9 \times 10^{-3}$ & $2.8$ \\
Z-I-Re20000 & $(1536,768,192)$ & $53$ & 20000 & 400 & 0.02 & $3.5 \times 10^{-2}$ & $2.6 \times 10^{-2}$ & $9.2 \times 10^{-3}$ & $2.8$ \\
Z-I-Re40000 & $(3072,1536,384)$ & $53$ & 40000 & 400 & 0.01 & $3.3 \times 10^{-2}$ & $2.3 \times 10^{-2}$ & $1.0 \times 10^{-2}$ & $2.4$ \\
\hline
Z-CL-192-a &  $(768,384,192)$ & $53$& - & 400 & - & $3.3\times 10^{-2}$ & $2.4\times 10^{-2}$ & $8.8\times 10^{-3}$ & $2.8$ \\ % case3 \\
Z-CL-192-b &  $(768,384,192)$ & $53$& - & 400 & - & $3.2 \times 10^{-2}$ & $2.3\times 10^{-2}$ & $9.3\times 10^{-3}$ & $2.5$\\ % case1\\
Z-CL-128-a &  $(512,256,128)$ & $53$& - & 400 & - & $3.5 \times 10^{-2}$ & $2.4\times 10^{-2}$ & $1.0\times 10^{-2}$ & $2.4$\\ % case6 \\
Z-CL-128-b &  $(512,256,128)$ & $53$& - & 400 & - & $3.4 \times 10^{-2}$ & $2.5\times 10^{-2}$ & $9.7\times 10^{-3}$ & $2.5$\\ % case4 \\
Z-CL-96-a &  $(384,192,96)$ &$53$ & - & 400 & - & $3.4 \times 10^{-2}$ & $2.5\times 10^{-2}$ & $9.2\times 10^{-3}$ & $2.8$\\ % case9 \\
Z-CL-96-b &  $(384,192,96)$ & $53$& - & 400 & - & $3.7 \times 10^{-2}$& $2.6\times 10^{-2}$ & $1.1\times 10^{-2}$ & $2.3$ \\ % case7 \\

\hline\hline
\end{tabular}
\caption{Same than table \ref{tab:runBy} but for SNOOPY runs. All runs have a box size of
  $(L_x,L_y,L_z)=(4L_z,4L_z,L_z)$.}
\label{tab:runInc}
\end{center}
\end{table*}

The whole set of runs discussed in the remaining of this paper is listed in
tables \ref{tab:runBy}, \ref{tab:runBz} and \ref{tab:runInc}, where
the first column provides the simulation labels. Runs starting with a
pure azimuthal (vertical) magnetic field are labelled with letter 
``Y'' (``Z''). Likewise, compressible (incompressible) simulations are
labelled with letter ``C'' (``I''). Large Eddy Simulations (LES) are
labelled either ``ILES'' or ``CL'' depending on the subgrid scale
model that is used (see section~\ref{sec:LES}). The remaining columns
in tables \ref{tab:runBy}, \ref{tab:runBz} and \ref{tab:runInc} give
the run resolution (col. 2) and duration $T_{Run}$ (col. 3), the
Reynolds number (col. 4), the magnetic
Reynolds number (col. 5), the magnetic Prandtl number (col. 6), and
$\alpha$ (col. 7), the sum of $\alpha_{Rey}$ (col. 8) and
$\alpha_{Max}$ (col. 9). The later are defined by the following
relations: 
\ba
\alpha_{Rey} = \left\{ 
\begin{array}{l l}
  \frac{\langle \rho \delta v_r \delta u_\phi\rangle}{P_0} \,& \quad \text{in compressible runs,}\\
   \frac{\langle\delta v_r \delta u_\phi\rangle}{(\Omega Lz)^2} & \quad \text{in incompressible runs,}\\ \end{array} \right.  \label{eq:alpharey}\\
\alpha_{Max} = \left\{ 
\begin{array}{l l}
 -\frac{\langle B_rB_\phi\rangle}{P_0} \,& \quad \text{in compressible runs,}\\
  -\frac{\langle B_rB_\phi\rangle}{\rho_0(\Omega Lz)^2} & \quad \text{in incompressible runs.}\\ \end{array} \right. \label{eq:alphamax}
\ea
Here $\langle.\rangle$ denotes an average over the shearing box volume and over
time and $\delta v$ is the velocity difference to the laminar sheared
flow. 
Except for the shortest runs (see for instance section~\ref{sec:GChallenge}),
the turbulent transport rates as measured by the $\alpha$ parameters
are time averaged over the last $60$ orbits of the models. The last
column finally gives the ratio between magnetic and hydrodynamic transport
rate.

\subsection{Direct Numerical Simulations (DNS) in the small Pm regime}
\label{sec:DNS}

In this section, we present the results of the resolved
simulations (meaning that kinematic viscosity and ohmic resistivity
are explicitly included in the calculation), focusing on the rate of
angular momentum transport and on the power spectra of the turbulent
flow. 

\subsubsection{A $1000^3$ MRI simulation}
\label{sec:GChallenge}

\begin{figure}
\begin{center}
\includegraphics[scale=0.4]{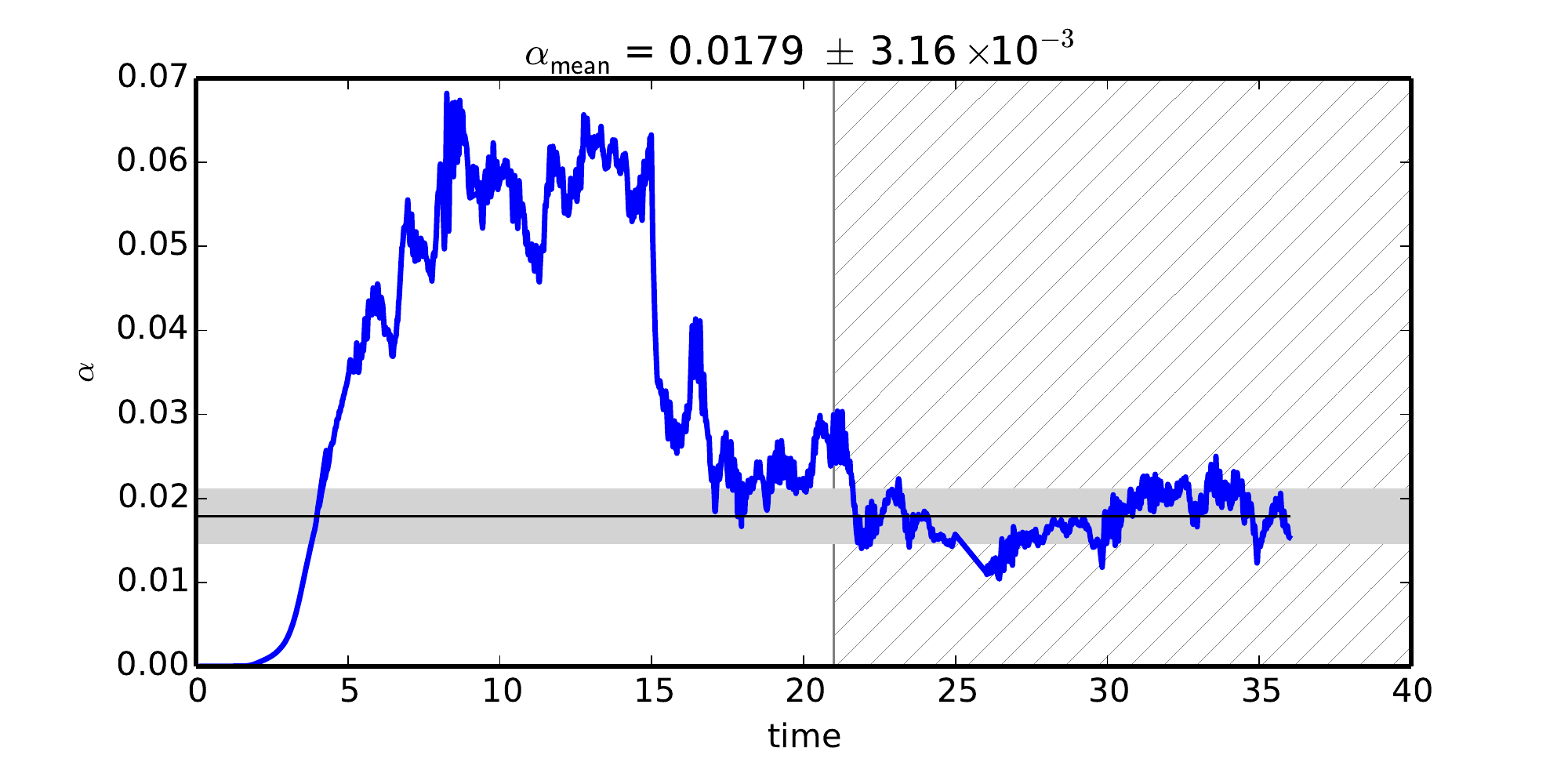}
\includegraphics[scale=0.41]{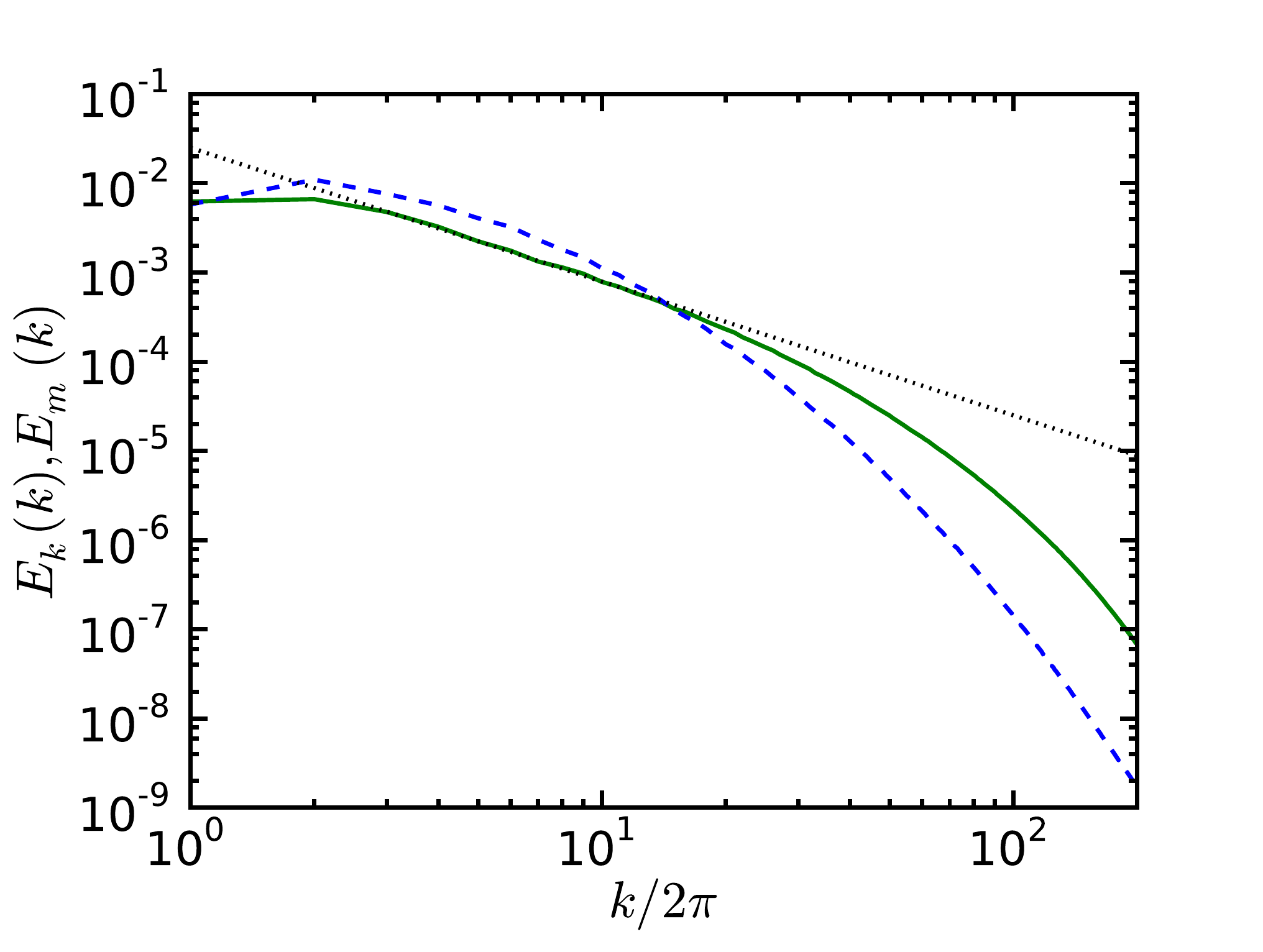}
\caption{Top panel: Time evolution of $\alpha$ in model Y-C-Re85000 for which $Pm=0.03$. The initial $15$
orbits were performed in the \emph{ideal MHD} regime, before including
explicit diffusion coefficients. The hatched region represents the
time interval over which the angular momentum transport is
averaged. Bottom panel: Y-C-Re85000 run, kinetic (solid green
line) and magnetic energy (dashed blue line) power spectrum.} 
\label{fig:turingAlpha}
\end{center}
\end{figure}

\begin{figure*}[!ht]
\begin{center}
\includegraphics[scale=0.32]{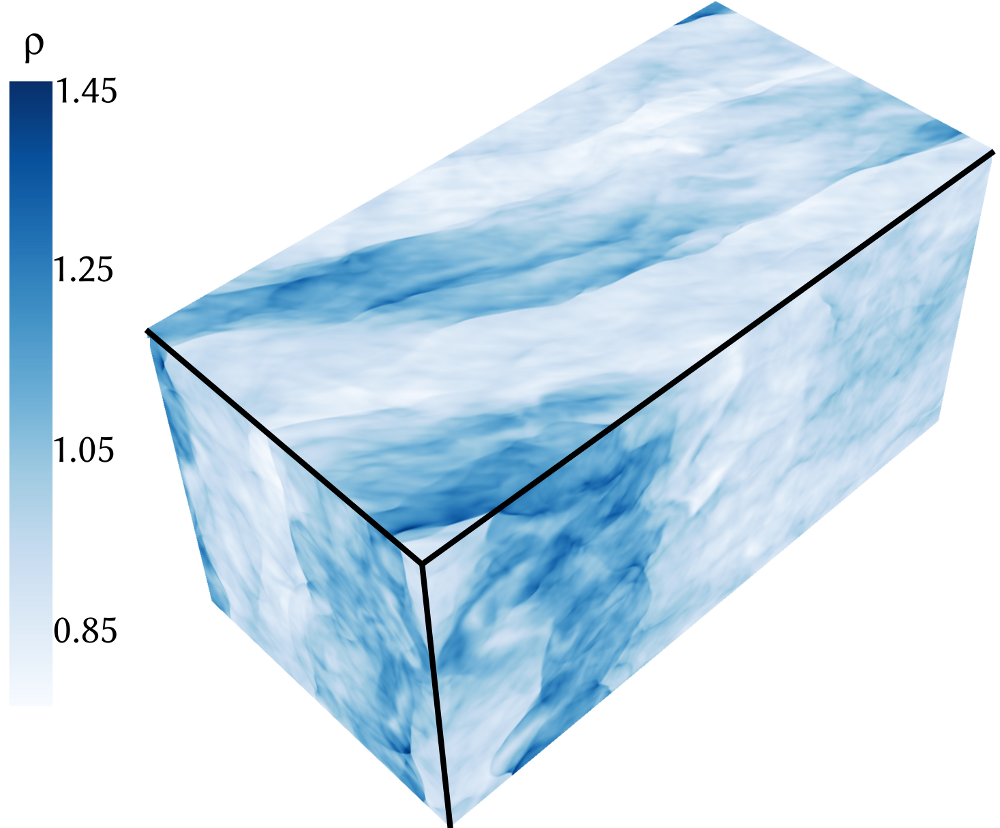}
\hspace{0.2cm}
\includegraphics[scale=0.32]{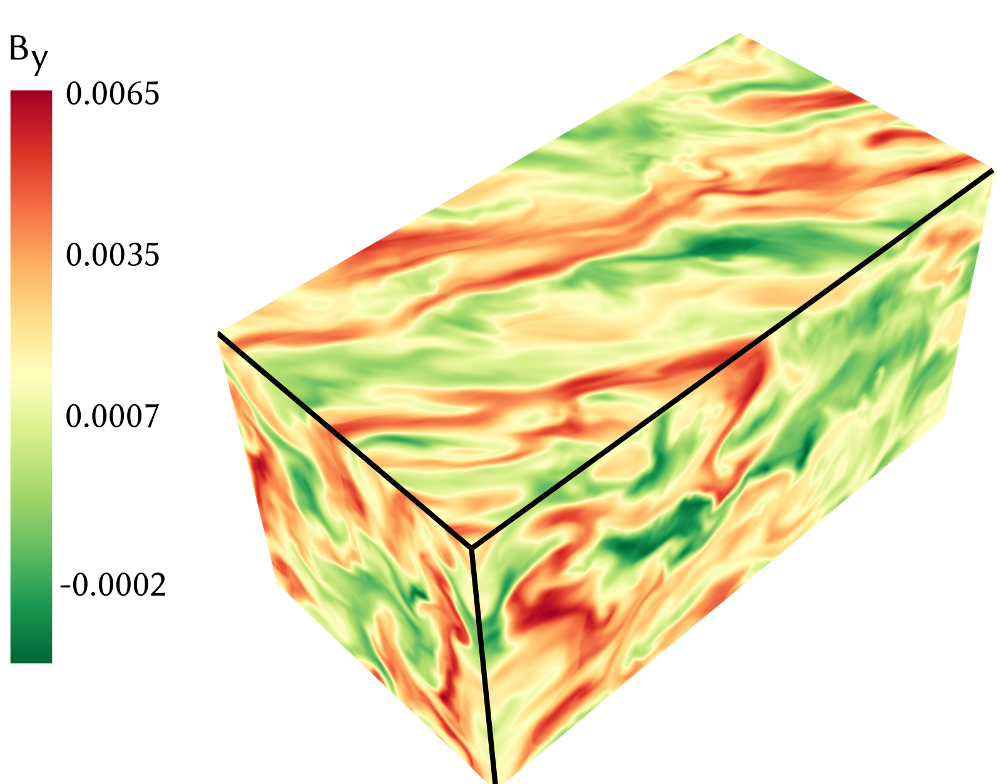}
\hspace{0.2cm}
\includegraphics[scale=0.32]{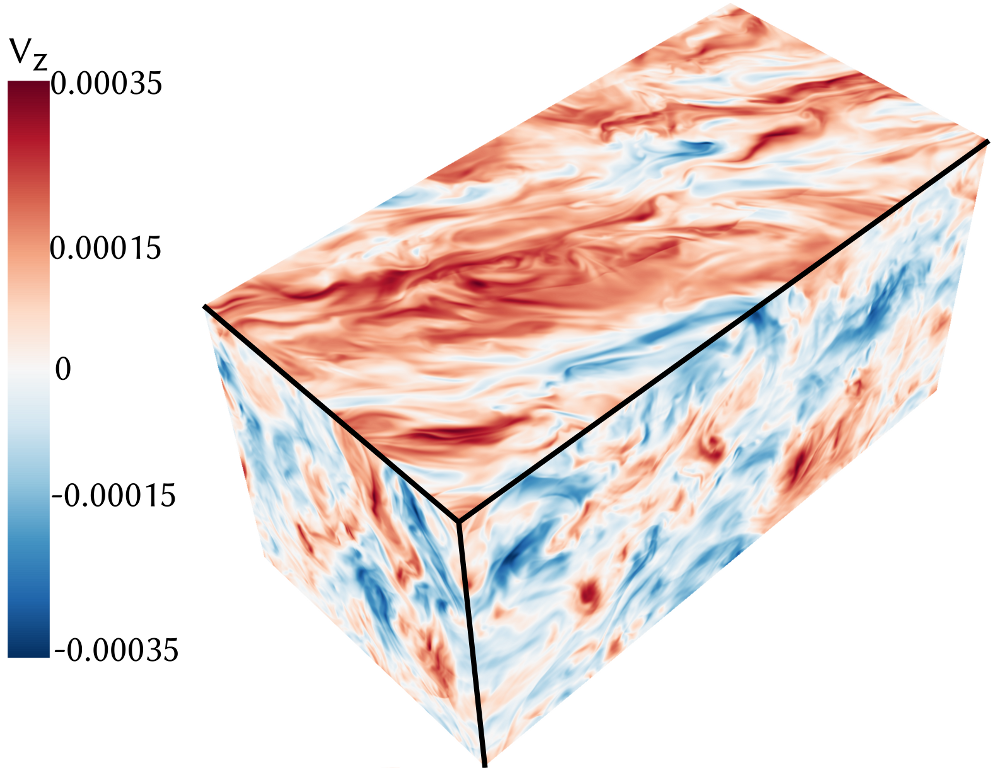}
\caption{From left to right, 3D snapshots of $\rho$, $B_y$ and $v_z$
in model Y-C-Re85000 at the end of the simulation.}
\label{fig:turingSnapshots}
\end{center}
\end{figure*}

Because of the tremendous computational cost that was
associated with that simulation, we start with a description of model
Y-C-Re85000, performed with RAMSES. In this model, $Re=85000$ and
$Pm=0.03$. 
This simulation was performed with a resolution
$(N_x,N_y,N_z)=(800,1600,800)$. Based on past experience
and published results of simulations using the same setup at lower
Reynolds number with a similar code \citep{SH09}, we are
confident that the small scales of the flow are properly accounted for
in that simulation. In the $x$ and $z$ directions, the number of cells  thus amounts to $800$ cells per disc scaleheight, which is the highest
resolution ever achieved of MRI--driven MHD turbulence with a 2nd
order compressible code. This gigantic simulation was run at the IDRIS
supercomputing center on the BlueGen-Q machine {\it Turing}, using
$32768$ cores. We found that the most efficient configuration was to
use $4$ threads per cores, meaning that a total of $131~072$ threads
(or, equivalently, MPI sub--domains) were used. We used approximately
$10$ hours of CPU time per timesteps. Running the simulation for over
$35$ orbits ($\sim 10^6$ timesteps) thus required about $10^7$ hours
of CPU time, corresponding to $300$ hours of wall clock time
(i.e. about two weeks). 

As described by \citet{SH09}, in the presence of a mean
azimuthal magnetic field, resistivity can prevent the linear
instability from transiting to a turbulent state when the simulation
starts from a laminar flow, even though it is found to remain
turbulent on long timescales when that linear phase is bypassed. We
thus adopted the method described by these authors and performed the run is two steps:
we first solved the ideal MHD equations (i.e. with vanishing viscosity
and resistivity parameters). The only dissipative effects are numerical
in origin during that part of the calculation. A turbulent state is
reached after $\sim 10$ orbits. The turbulent 
transport as measured by $\alpha$ displays fluctuations around a well
defined mean value of about $6 \times 10^{-2}$ for the next few orbits
(see figure~\ref{fig:turingAlpha}, top panel). At $t=15$, we
restarted the simulation for an additional $20$ orbits with dissipation
coefficients such that $Re=85000$ and $Pm=0.03$. As seen on the top
panel of figure~\ref{fig:turingAlpha}, the turbulence amplitude is
quickly modified by the presence of those dissipation coefficients
and reaches a new steady state after a few additional
orbits\footnote{Note that all the runs with a mean azimuthal 
  magnetic field are executed using the same procedure. However,
  except for model Y-C-Re85000 for which the computational cost is
  considerable, they have usually been run for $100$ orbits to obtain
  a more precise measure of $\alpha$.}. The nature of the flow is 
illustrated by a series of snapshots of the last time step of the run in
figure~\ref{fig:turingSnapshots}. On the density plot one can recognize
large scales density waves and low amplitude shocks. The turbulent
magnetic field is dominated by large scale structures. On the
contrary, the velocity snapshot shows both large and small scales
structures, as expected given the very small $Pm$ used here.  

We next averaged the transport rate after the system has reached a
quasi steady state and until the end of the simulation. This period
corresponds to $t>21$ orbits and is hatched on
figure~\ref{fig:turingAlpha} (top panel). The value of 
$\alpha_{Rey}$, $\alpha_{Max}$ and $\alpha$ we obtained are $4.6
\times 10^{-3}$, $1.3 \times 10^{-2}$ and $1.8 \times 
10^{-2}$, respectively (see also Table~\ref{tab:runBy}). This $\alpha$ value 
is comparable to the one reported by \citet{SH09} for
their most resolved run, for which $Pm=0.25$, $Rm=3200$ and $\beta=450$. The Maxwell and 
Reynolds stresses display a ratio of $\sim 3$ that is typical of such
simulations. There is of course a degree of arbitrariness
regarding the exact time at which we decide that the system has
reached ``a quasi steady state'' and start averaging. We have checked
that the statistics we consider do not vary significantly when making
modest changes to the averaging period. For example, we find
$\alpha=1.9\times 10^{-2}$ and $\alpha=1.8\times 10^{-2}$ when averaging over the periods $t>18$ and
$t>25$ orbits, respectively. This is only a variation by $7\%$ and
give an idea of the uncertainty associated with that measurement.\\

\begin{figure*}[!ht]
\begin{center}
\includegraphics[scale=0.75]{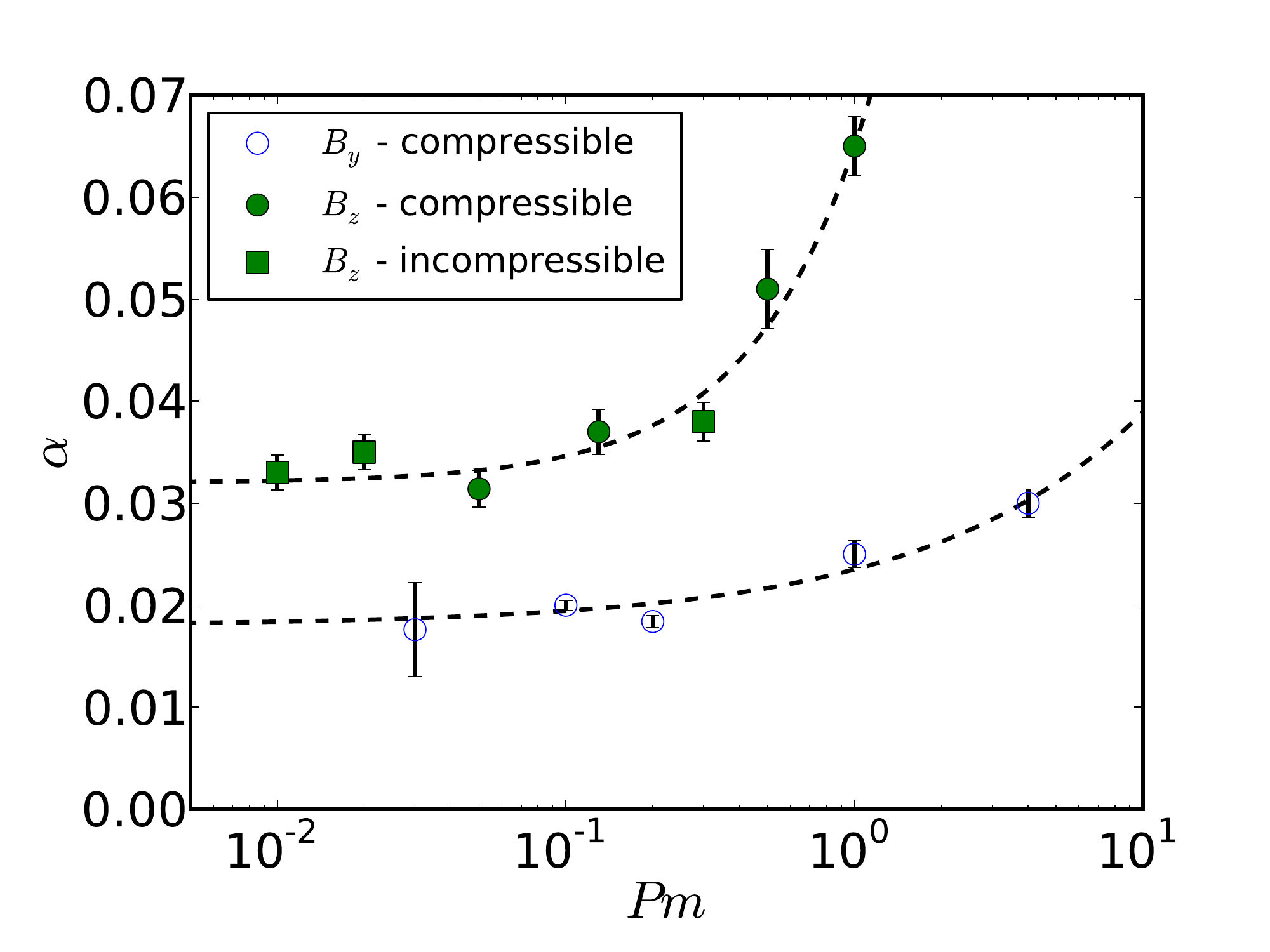}
\caption{Mean value of the angular momentum transport (measured by
means of $\alpha$) for the mean toroidal field simulations ({\it blue
circles}), the mean vertical field simulations performed with RAMSES
({\it green filled circles}) and with SNOOPY ({\it green filled squares}). 
The dashed lines are power-law functions
that approximately describe the data (see text).} 
\label{fig:convAlpha}
\end{center}
\end{figure*}

To obtain a more accurate view of the energy budget at each scales, we
also show on figure~\ref{fig:turingAlpha} (bottom panel) the time
averaged power spectra. Because of the 
shearing box boundary conditions, we consider time dependent unsheared
wave vector $\vec k$ and the shell filter decomposition of the
physical variables \citep{HGB95}. Being spherically symmetric, we note that
this method filters out the information about the flow anisotropy which
is known to be significant for MRI-driven turbulence (see \citealt{MP15}, Figure 3. of \citealt{LL11} and section \ref{sec:CL}). For each wavenumbers, the
kinetic and magnetic energy are then given by
\begin{eqnarray}
E_K(k) &=& \overline{v^2_k(k)/2} \\
E_M(k) &=&\overline{\vec B^2_k(k)/2}  \, ,
\end{eqnarray}
respectively, where the bar denotes an average over time. Magnetic energy dominates at large scales ($k'<15$, where we
have defined $k'=k/2\pi$ so that it corresponds to scales $l'>H/15$)
while kinetic energy dominates at smaller 
scales. This is a signature of the small $Pm$ value of the simulation,
which implies that the resistive dissipation length is larger than the
viscous dissipation length \citep{SCT04}. At small
scales, the motions essentially correspond to 
hydrodynamic turbulence associated with a forward cascade
\citep{LL11}. We will exploit that scale separation in
section~\ref{sec:LES} when designing sub-grid scale models for the
hydrodynamic part of the flow. The kinetic energy power spectrum
follows a power law with exponent $-3/2$ over the range $2<k'<20$,
thus covering one order in magnitude in spatial scales. The
same exponent has been reported recently in other high resolution
simulations of MRI--driven MHD turbulence both in the dynamo regime
\citep{F10} and in the presence of a net vertical field \citep{LL11},
suggesting it is a general feature of the flow. We note that
homogeneous forced MHD turbulence also displays the same exponent
\citep{MCB08}, although the result is still debated
\citep{BE14}. The surprising result here is that
the magnetic energy does not show any obvious signature of a power--law
regime, contrary to the case of homogeneous and driven MHD turbulence,
while still being comparable in magnitude with $E_K$. More
work is needed to clarify the reason for that discrepancy and to
better understand the origin of the $-3/2$ exponent that is obtained
in the case of MRI--driven MHD turbulence. 

\subsubsection{Angular momentum transport}
\label{sec:angmom}

We plot on figure~\ref{fig:convAlpha} the total averaged stress
$\alpha$ for all the resolved simulations we performed. Error bars
$\sigma_{\alpha}$ are estimated with the method presented in
\citet{LL10}. In general, we find $\sigma_{\alpha} \sim 5 \times
10^{-3}$ for the vertical field model (thus giving
$\sigma_{\alpha}/\alpha \sim 5 \%$). This is consistent with the
estimate of \citet{LL10}. The error estimate is smaller in the
azimuthal field case, for which we obtained $\sigma_{\alpha} \sim
10^{-3}$ which corresponds to $\sigma_{\alpha}/\alpha \sim 1 \%$. We note
that we could not apply the same method for model 'Y-C-Re85000'
because of the short duration of the integration in that case. A
simple standard deviation is thus plotted instead and explains the
larger error bar for that particular model.  

In agreement with previous results (see section~\ref{sec:intro}) we
find that $\alpha$ increases with Pm. However, the main result of
figure~\ref{fig:convAlpha} (and the main result of the paper) is that
there is now convincing evidence of the convergence of the angular
momentum transport rate at small $Pm$ toward a well defined, nonzero
value. This is the case for both field geometry. 

\paragraph{The azimuthal field case:} In this case, and with constant
$Rm$ and $\beta$, we find that a fit to the data is given by the formula 
\begin{equation}
\alpha_{By}=\alpha_{Bymin}+C_y Pm^{0.58} \, ,
\end{equation}
with the values\footnote{Note that the asymptotic transport values $\alpha _{Bymin}$ and $\alpha_{Bzmin}$ depend in principle on $Rm$ and $\beta$. See the discussion in section \ref{sec:conclusion}.} of $\alpha_{Bymin}=1.8 \times 10^{-2}$ and $C_y=5.5 \times
10^{-3}$, respectively. This means that a good estimate of $\alpha$
is already obtained at $Pm=0.2$ for which a resolution of $128$ cells
per unit length is sufficient. Indeed, we obtained
$\alpha=1.8 \times 10^{-2}$ which is equal to the asymptotic value of
$\alpha$ at vanishingly low $Pm$.

\begin{figure*}[!ht]
\begin{center}
\includegraphics[scale=0.45]{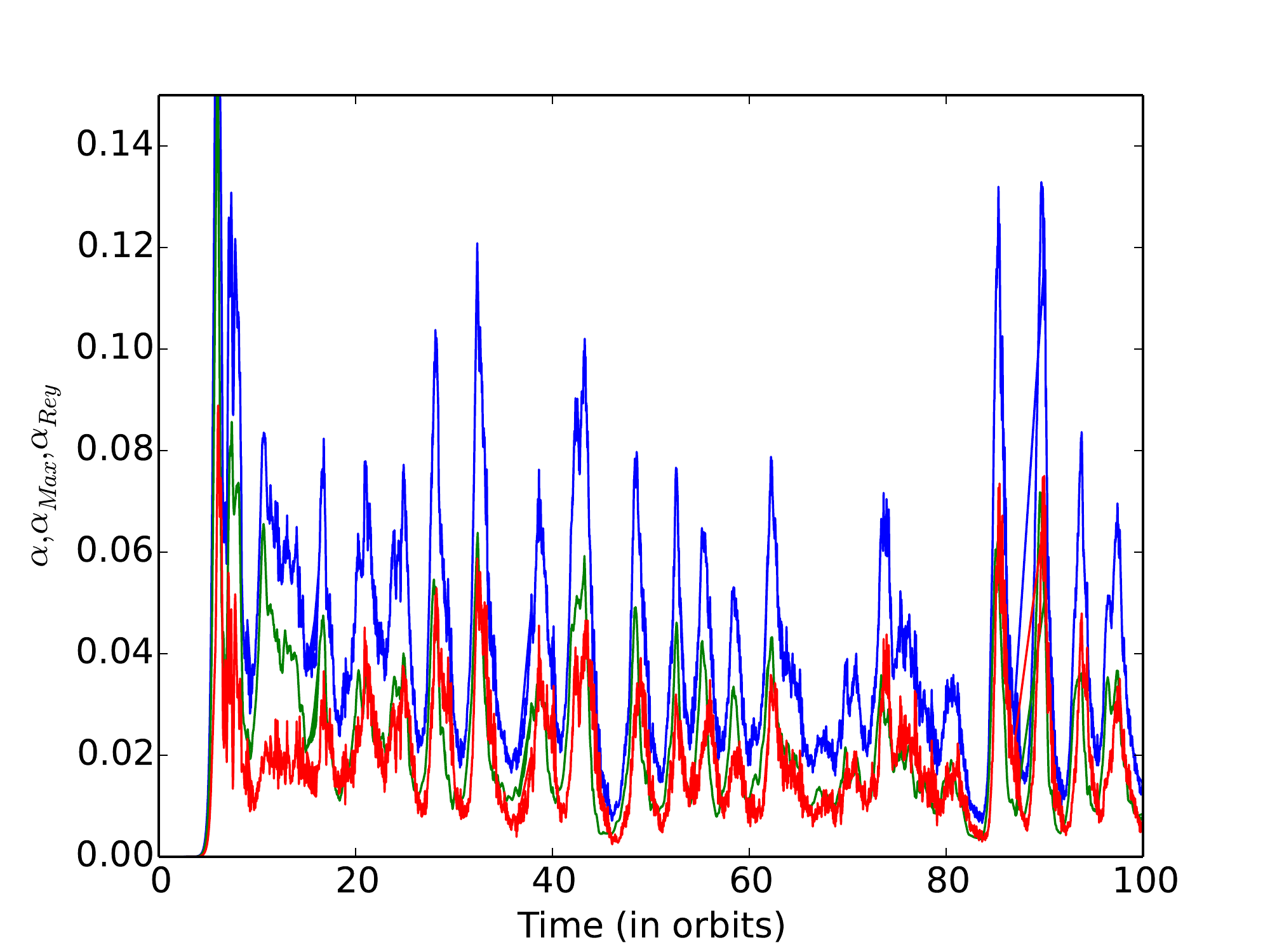}
\includegraphics[scale=0.45]{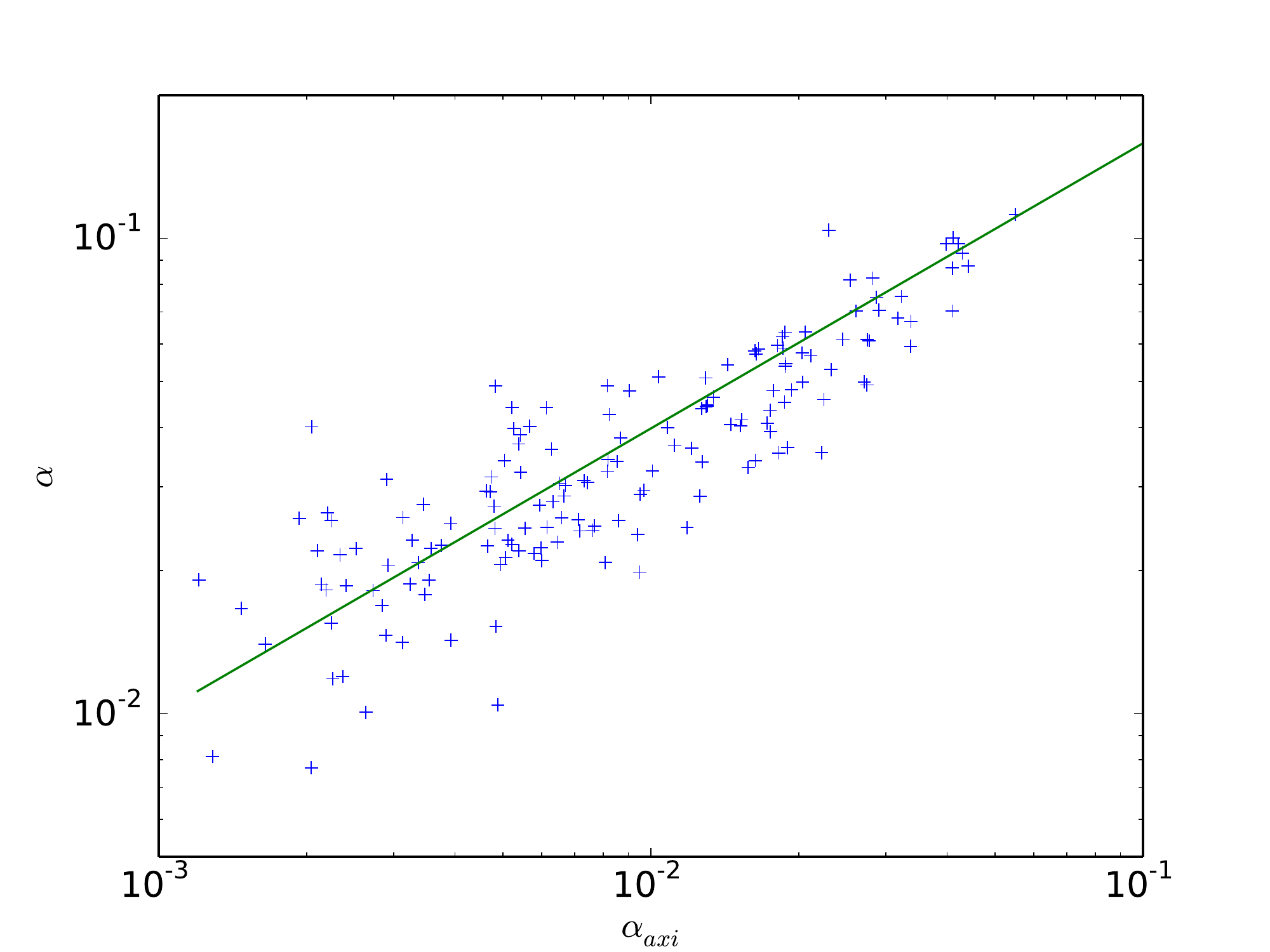}
\caption{Left panel: Time history of $\alpha$ ({\it blue curve}), $\alpha_{Max}$
  ({\it red curve}) and $\alpha_{Rey}$ ({\it green curve}) in model
  Z-C-Re3000. Right panel: scatter plot showing the total angular
  momentum transport rate $\alpha$ as a function of $\alpha_{axi}$
  which measures the angular momentum transport due to $k_y'=0$ and $k_z'=1$
  modes for $120$ dumps evenly spaced over model Z-C-Re3000. The green
curve plots an approximate fit to the data (see text for details).} 
\label{fig:alphaBursts}
\end{center}
\end{figure*}

\paragraph{The vertical field case:} Here, we find a somewhat larger
transport coefficient that can be fitted by the relation 
\begin{equation}
\alpha_{Bz}=\alpha_{Bzmin}+C_z Pm^{1.1}
\end{equation}
with $\alpha_{Bzmin}=3.2 \times 10^{-2}$ and $C_z=3.3 \times
10^{-2}$, with fixed $Rm$ and $\beta$.
Moreover, there is good agreement between the compressible
and incompressible approach in this vertical field case: for both flow type the simulations converge at small
$Pm$ toward the same $\alpha$. As for the
azimuthal field case, a good estimate of the asymptotic rate of
angular momentum transport is already obtained for $Pm=0.13$ (using a
resolution of $128$ cells per unit length), for which we found
$\alpha=3.7 \times 10^{-2}$, i.e. a value that differs by about
$15\%$ from the asymptotic limit.

As already noted (section \ref{sec:model_params}), in the presence of
a vertical magnetic field we find that the time history of $\alpha$
displays significant variability. This is illustrated on
figure~\ref{fig:alphaBursts} (left panel) for the particular case of
model Z-C-Re3000 ($Pm=0.13$): while the time averaged transport
rate amounts to $\alpha=3.7 \times 10^{-2}$ in that case, there are
numerous peaks during which it reaches values as high as $0.1$
that occur with a typical period of about $5$
to $10$ orbits. Such bursts are not unheard of in unstratified
shearing boxes with a mean vertical field
\citep{BMC08,LLB09} and have been attributed to recurrent
`channel-like' modes associated with the MRI. For the set of parameters we
have considered here, the time history of $\alpha$ suggests that they
contribute significantly to the turbulent transport. In an attempt to
quantify that contribution, we have calculated for model Z-C-Re3000 
the value $\alpha_{axi}$ of the transport that is due to axisymmetric
`channel-like' modes (for which $k_y'=0$ and $k_z'=1$) for $120$ dumps evenly
spaced between $t=40$ and $t=100$. The mean value of
$\alpha_{axi}$, averaged over all the dumps of the simulation,
amounts to $1.2 \times 10^{-2}$: this means that
axisymmetric modes with $k_z'=1$ accounts for $\sim 30 \%$ of the angular
momentum transport. In agreement with the results of \citet{LL10},
angular momentum transport is dominated on average by non-axisymmetric
modes even if the contribution of `channel-like' modes is
significant. The scatter plot showing the relation between $\alpha$ 
and $\alpha_{axi}$ for those $120$ dumps is shown on the right panel of
figure~\ref{fig:alphaBursts}, along with an indicative fit of the data
given by 
\begin{equation}
\alpha=\alpha_0 \left( \frac{\alpha_{axi}}{\alpha_{axi0}}
\right)^{0.6} \, ,
\label{eq:fitScatter}
\end{equation}
with $\alpha_0=10^{-2}$ and $\alpha_{axi0}=10^{-3}$. The positive
correlation between $\alpha$ and $\alpha_{axi}$ indicates that the
relative contribution of the transport mediated by axisymmetric modes
with $k_z'=1$ increases during the bursts of activity: for example,
$\alpha_{axi}$ amounts to only about $10 \%$ of the transport when
$\alpha=10^{-2}$ but, as indicated by Eq.~(\ref{eq:fitScatter}), can
contribute for up to $50 \%$ of the turbulent activity when
$\alpha=0.1$. These variations are consistent with the results of
\citet{LLB09} and with the idea that the bursts seen on
figure~\ref{fig:alphaBursts} are due to large scale `channel-like' modes
\citep{BMC08}. We have repeated the same 
analysis for all the models and we have found a weak dependance of the
relative fraction of axisymmetric transport with $Pm$:
$\alpha_{axi}/\alpha=0.39$, $0.34$, $0.32$ and
$0.29$, respectively for $Pm=1$, $0.5$, $0.13$ and $0.05$ and similar
results for the incompressible runs with $\alpha_{axi}/\alpha=0.38$, $0.35$, respectively for $Pm=0.02$ and $0.01$,

 \begin{figure}
 \begin{center}
 \includegraphics[scale=0.4]{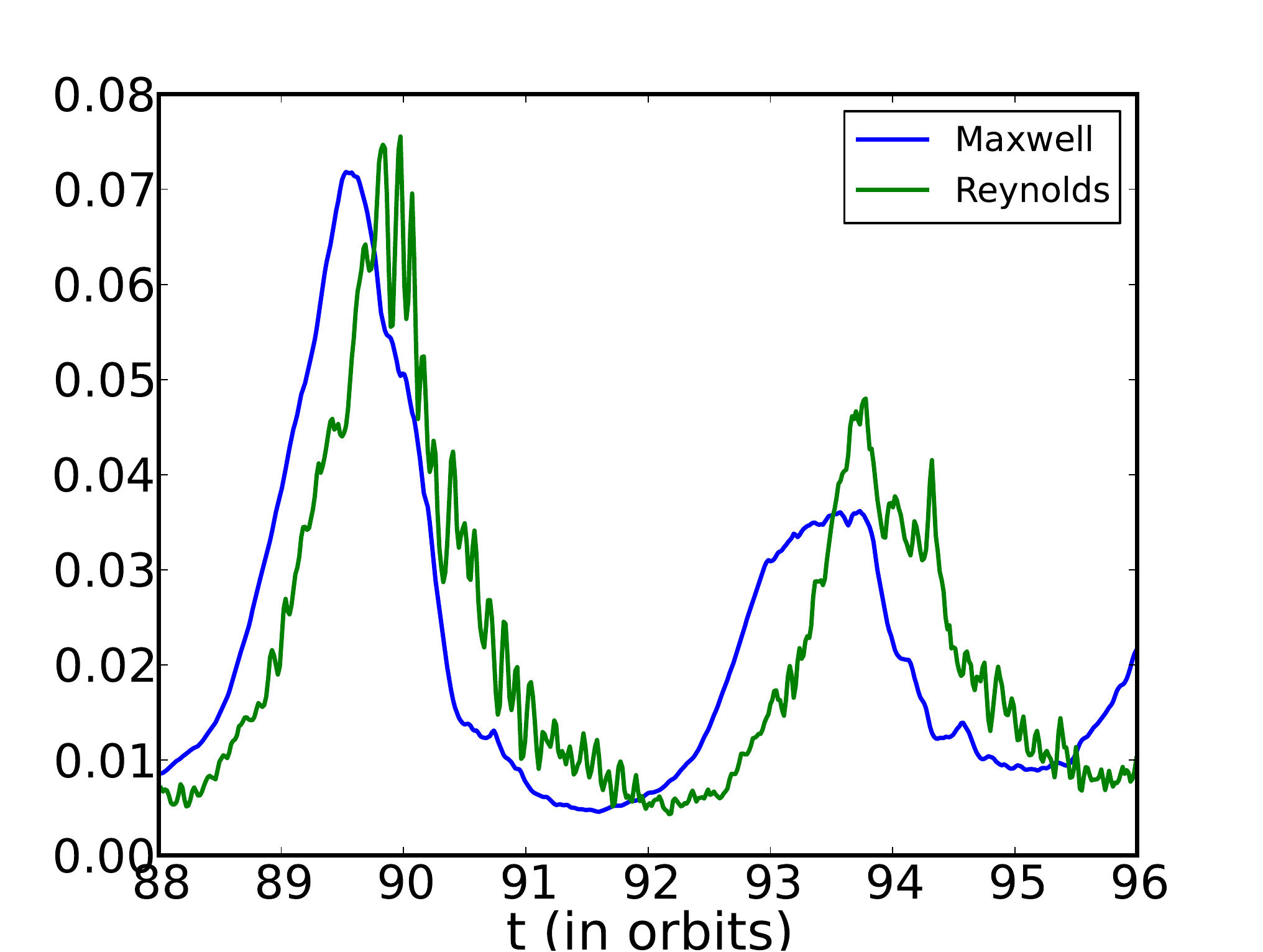}
 \caption{Time history of $\alpha_{Max}$
  ({\it blue curve}) and $\alpha_{Rey}$ ({\it green curve}) in model
  Z-C-Re3000.}
 \label{fig:alphaZoom}
 \end{center}
 \end{figure}

It is also noteworthy that the angular momentum transport in the presence
of a vertical field is equally due to Maxwell and Reynolds stresses in the
compressible simulation, whereas the classical ratio of approximately $3$ is
obtained in an incompressible fluid or with an azimuthal field configuration for the specific values of $\beta$ and $Rm$ chosen in our investigation.
The exact value of Maxwell to Reynolds stress ratio ($R$) are given in the last 
column of Tables \ref{tab:runBy},\ref{tab:runBz}, and \ref{tab:runInc}. In an attempt to
have a better understanding of the relative contribution of Maxwell and
Reynolds stresses, we plot on figure \ref{fig:alphaZoom} their time history
over two burst cycles for the particular case of
model Z-C-Re3000. 
This plot reveals that the Reynolds stress is
larger than the Maxwell stress during the decaying part of the burst when the
`channel-like' modes (with $k_y=0$ and $k_z=1$) are destroyed by the non-linear turbulent dynamics.
The computation of the contribution of the axisymmetric modes shows that
the Maxwell transport is dominated by `channel-like' modes whereas
the Reynolds transport is due to non-axisymmetric modes.
Because of the difference between the incompressible and compressible
runs, we further speculate that such a destruction is associated with the excitation 
of compressible modes such as density waves.
However this detailed study, not directly related to the angular momentum transport rate, is beyond the scope of this paper.

%%%%%%%%%%%%%%%%%%%%%%%%%%%%%
%%%%%%%%%%LES
\subsection{Large Eddy Simulations in the small Pm limit}
\label{sec:LES}

\begin{figure}
\begin{center}
\includegraphics[scale=0.45]{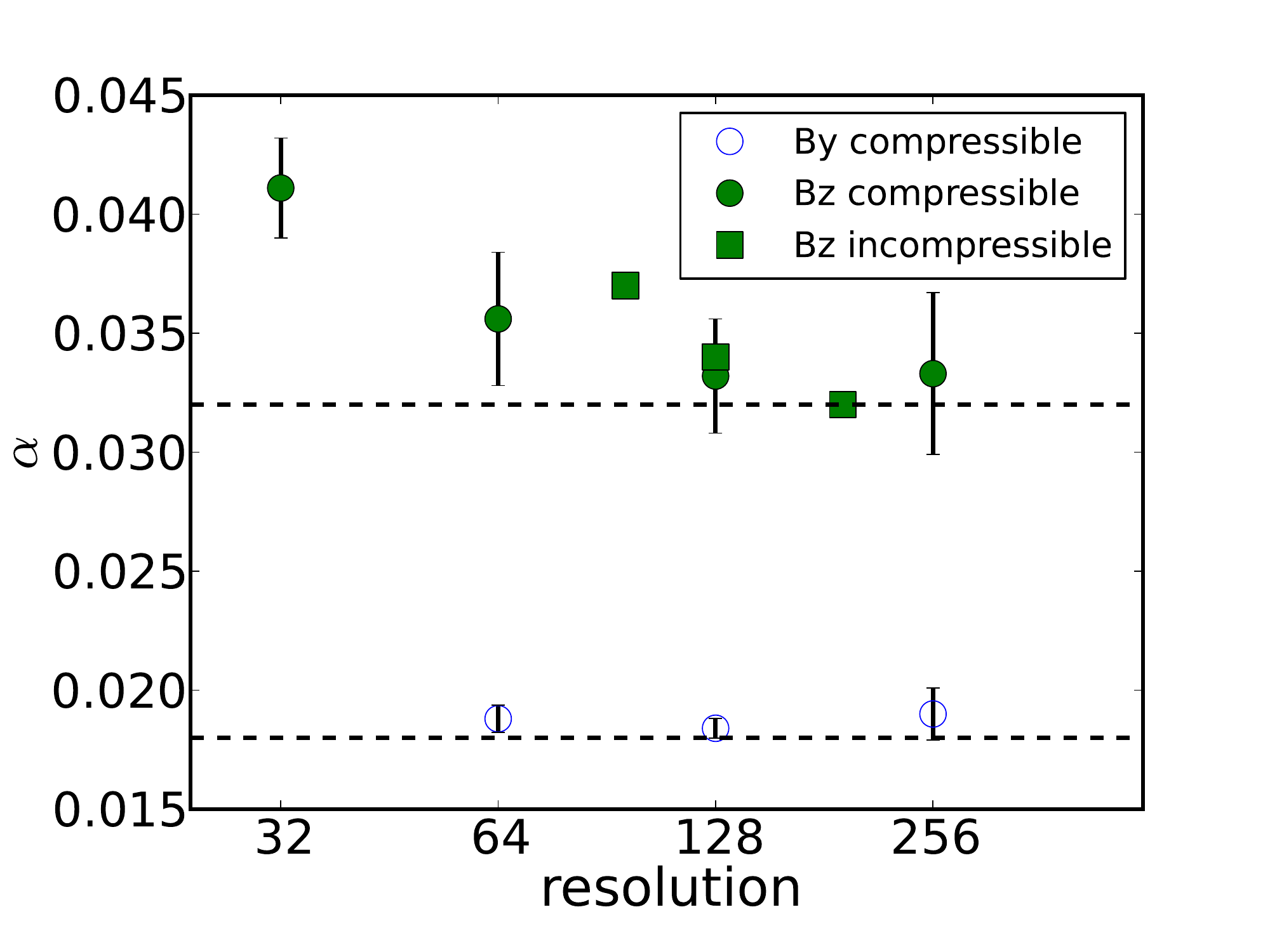}
\caption{Mean value of the angular momentum transport (measured by
means of $\alpha$) for the mean toroidal field LES simulation (implicit LES {\it blue
circles}), the mean vertical field LES simulation performed with RAMSES
(implicit LES, {\it green filled circles}) and with SNOOPY (Chollet-Lesieur LES, {\it green filled squares}).
The resolution is given in number of cells per scaleheight. The dashed lines represents the 
asymptotic limits obtained with DNS.} 
\label{fig:alphaLES}
\end{center}
\end{figure}

The previous results have shown that angular momentum transport
converges toward a well defined limit at small $P_m$. However, such
simulations are very computationally expensive. Here, we investigate
the possibility to 
use a sub-grid scale model instead of standard kinematic
viscosity. The aim is to reduce the computational cost of the
simulations without compromising the physics one may want to
consider, such as the accretion rate or the turbulent power spectra. 
  
Historically, the study of small $Pm$ flows has mostly relied on
simulations using hyperviscosity such as geodynamo models \citep{GR95}
or small scale dynamo theory \citep{SIC07}. However, hyperviscosity is known
to produce numerous artefacts in hydrodynamic turbulence, such
as spectral bottlenecks, reduced intermittency and spurious isotropisation
\citep{FKP08}. For this reason, we instead focus on Large Eddy Simulations (LES) in which
the dissipation adapts dynamically to the turbulent cascade initiated by the large scales, 
thereby limiting the artefacts commonly due to hyperviscosity.

LES are routinely used in industrial applications to model flows at
high $Re$. However, their MHD counterpart are not widespread, the reason
being the higher complexity of the MHD turbulent cascade compared to the purely
hydrodynamic cascade. Several subgrid-MHD models have been discussed in the literature including
\cite{S63} type models \citep{Y87} which can be used in finite difference/volume codes
and \cite{CL81} type models \citep{BPP08} targeted to spectral methods. However, most of these 
models are still under development and their applicability to $Pm\ll 1$ flows is yet to be proven.

In this work we have chosen to use well-known hydrodynamical subgrid models to treat the kinetic turbulent cascade only, 
leaving the induction equation with standard Ohmic resistivity. This approach is valid provided that the subgrid model is introduced at a scale much
smaller than the resistive scale, so that the energy cascade is mostly hydrodynamical. Moreover, it has the advantage of using well-tested
subgrid models which are reasonably simple to implement numerically. This type of methods has already
been used to study Taylor-Green flows \citep{PPP04} and dynamo action \citep{PMM05} in the limit $Pm\rightarrow 0$. We therefore
reproduce this approach using very similar tools in the MRI turbulence context.

%###############Chollet-Lesieur
\subsubsection{Chollet-Lesieur model in incompressible simulations}\label{sec:CL}

\begin{figure}
\begin{center}
\includegraphics[width=0.95\linewidth]{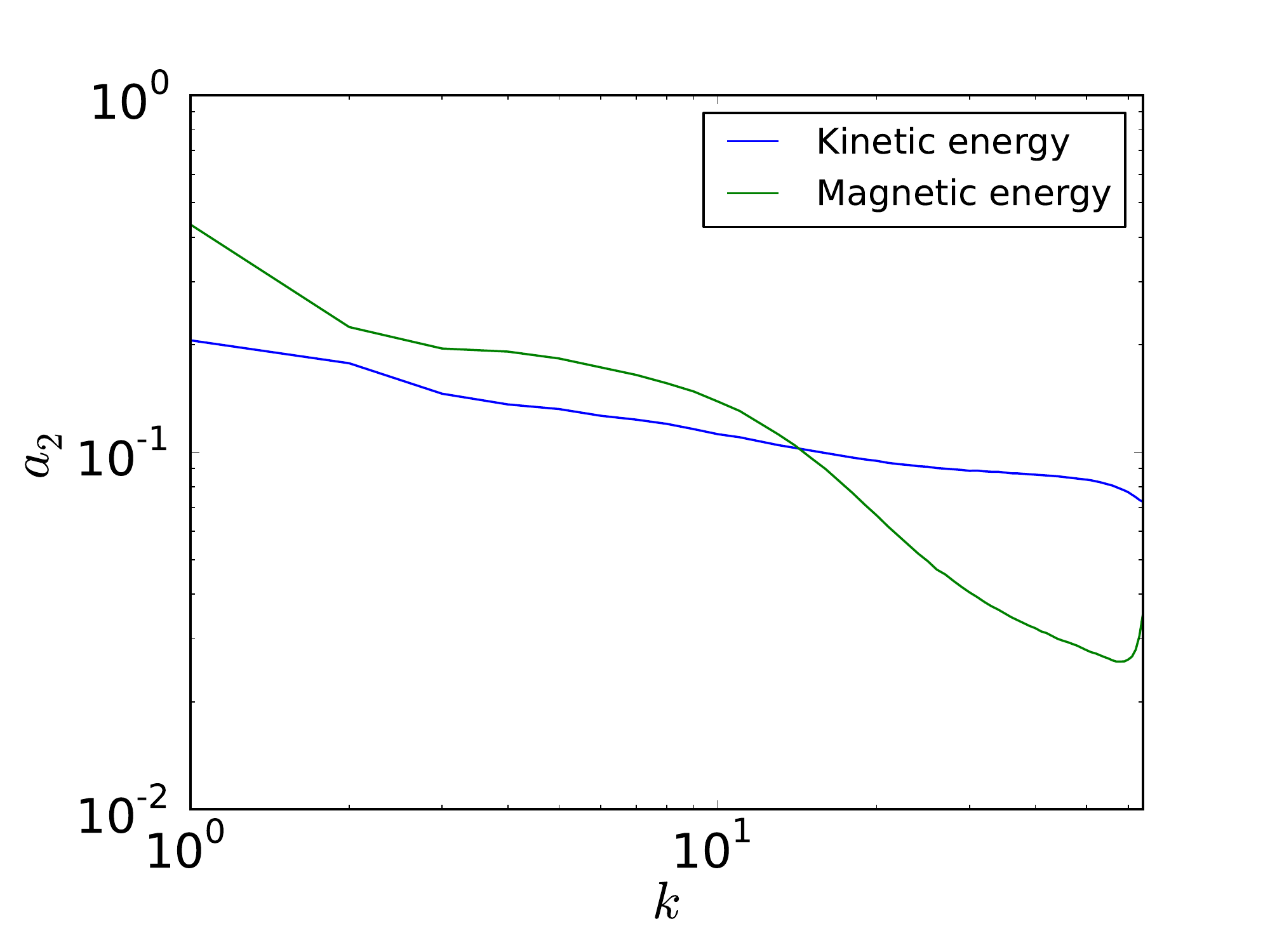}
\caption{Estimate of the anisotropy at each scale. For each wavenumber, the energy is decomposed into spherical harmonics; the anisotropy is estimated from the sum of the second order coefficients normalized by the isotropic coefficient.}
\label{fig:anisotropy}
\end{center}
\end{figure}

Since our incompressible simulations are done with a spectral code, we have used the spectral subgrid model of \cite{CL81} (CL) to perform our incompressible LES. This model replaces the standard viscosity $\nu$ by the following expression in Fourier space:
\begin{equation}
\nu_\mathrm{CL}(k)=\Bigg(\frac{E(k_c)}{k_c}\Bigg)^{1/2}\Big(0.267+9.21\exp[-3.03k_c/k]\Big)
\end{equation}
where $k_c$ is the cutoff scale and $E(k_c)$ is the kinetic energy at the cutoff scale\footnote{Note that several forms for the CL viscosity may be found in the literature since this expression is a fit to numerical EDQNM (Eddy-Damped Quasi-Normal Markovian) calculations. Our expression comes from the asymptotic viscosity of \cite{CL81} and a fit close to $k_c$ of \cite{S06}.} . This expression encloses long range nonlinear interactions via a constant viscosity at $k\ll k_c$ and a cusp due to local energy transfers close to the cutoff scale $k_c$. Interestingly, $k_c$ is the only free parameter of this model. The amount of viscosity itself is automatically adjusted according to the amount of energy at the cutoff scale. It should be emphasized that this expression is only valid for 3D homogeneous and isotropic Kolmogorov turbulence. In principle, it is therefore not applicable to turbulent shear flows found in shearing box models
since such flows are anisotropic (see section \ref{sec:GChallenge}). As a first attempt to quantify that anisotropy, we proceed as follows. We define a decomposition in spherical harmonics\footnote{This use of spherical harmonics to estimate the anisotropy of turbulence is a standard procedure to study sheared flows (e.g.\  \citealt{BV01}).} such that
\be
E_K(\bm{k})=\frac{1}{2}|\tilde{v}(\bm{k})|^2=\sum_{jm}c_j^m(|\bm{k}|)Y_{jm}(\hat{k}),
\ee
and similarly for the magnetic energy.
We focus on the $l=2$ coefficients of that decomposition and define:
\be
a_2=\sum_{m=-2}^{m=2} |c_2^m|^2/|c_0|^2.
\ee
$a_2$ takes significant values when there are strong variations of the energy over the shell,
or, in other words, when the flow displays anisotropy at that scale. As shown in figure \ref{fig:anisotropy}, it decreases with k but is still high at small scales for the kinetic energy. Nevertheless we will see that the Chollet-Lesieur model gives satisfactory results.

\begin{figure}
\begin{center}
\includegraphics[width=0.95\linewidth]{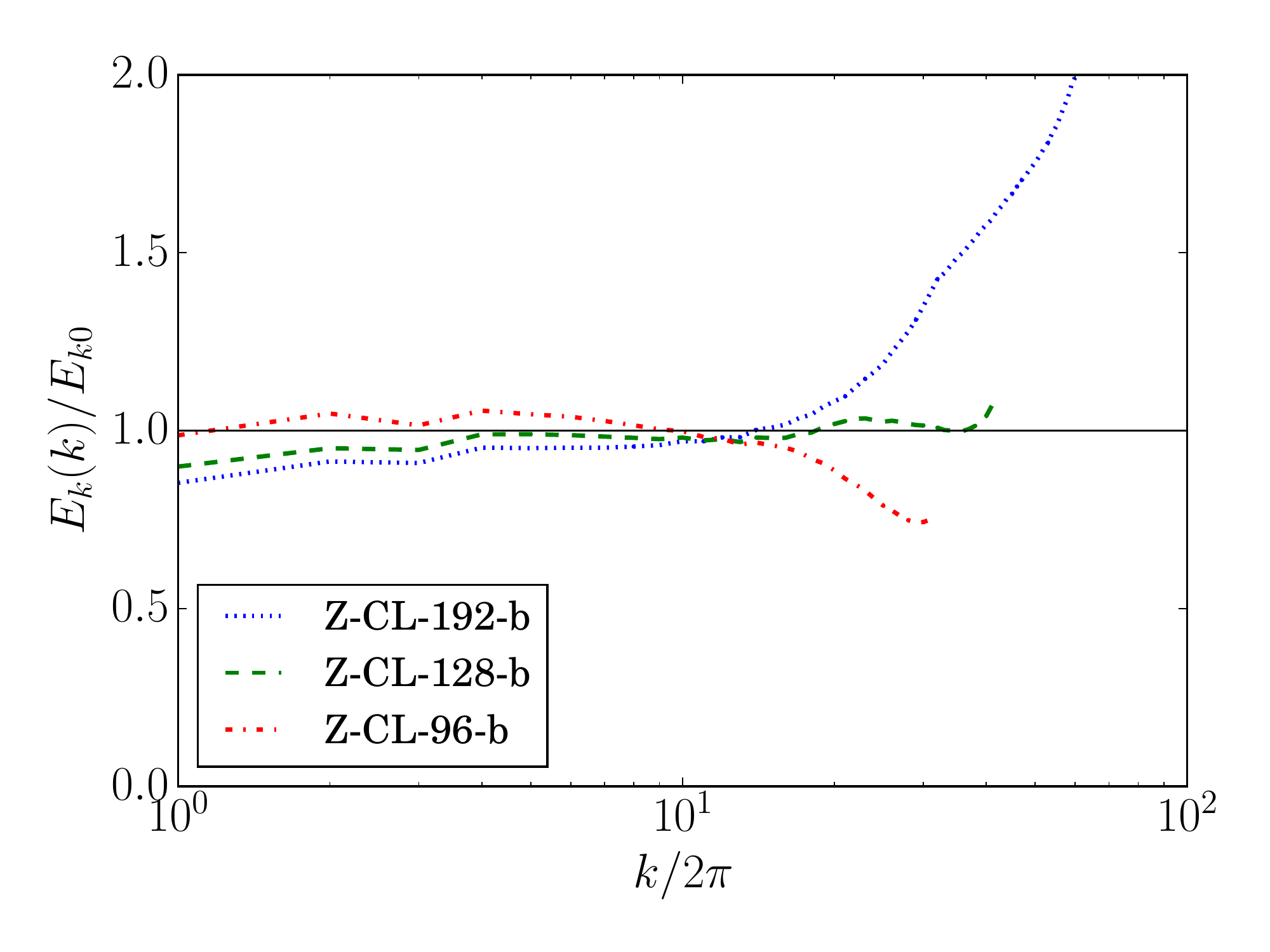}
\includegraphics[width=0.95\linewidth]{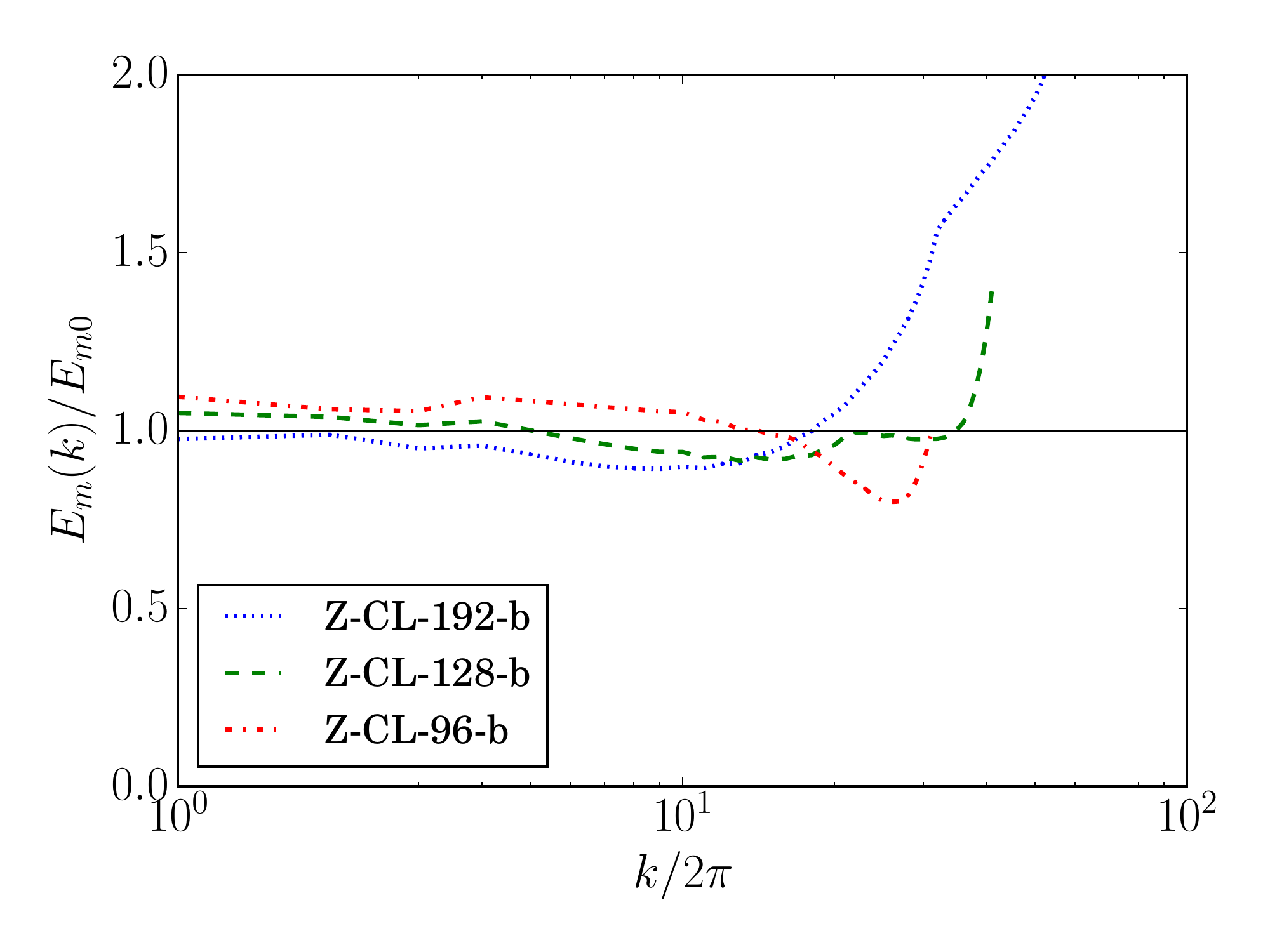}
\caption{Ratio of power spectra of kinetic energy (top) and magnetic energy (bottom) obtained with a Chollet-Lesieur subgrid model with $k_c/k_\mathrm{max}=1$ and full DNS calculation at Re=40000.}
\label{fig:spectra_CL}
\end{center}
\end{figure}

We have performed several simulations using CL subgrid model, varying $k_c$ and the resolution (runs Z-CL-XXX-X). Models labelled $a$ have $k_c/k_\mathrm{max}=0.75$ and models labelled $b$ have $k_c/k_\mathrm{max}=1$ where $k_\mathrm{max}$ is the maximum accessible wavenumber of the simulation (Table \ref{tab:runInc}). As plotted on figure \ref{fig:alphaLES} (green squares), all of our models recover the statistical results of DNS with a reduced resolution. These encouraging results are confirmed by turbulent spectra compared to DNS simulation (figure~\ref{fig:spectra_CL}). We find that both kinetic and magnetic spectra agree to less than 10\% down to $k\sim 20$ with the highest resolution DNS Z-I-Re40000. We also find that simulations with $k/k_c=0.75$ exhibit the same convergence properties, at least for the resolution we studied. Therefore, the cutoff scale does not seem to have much impact on these results, provided it is below the resistive dissipation scale. These results demonstrate that CL models can be used efficiently to study low $Pm$ flows, with a gain in resolution of at least a factor 2 associated with a gain in computation time of at least a factor 20.

%###############ILES
\subsubsection{Implicit LES in compressible simulations}

\begin{figure}
\begin{center}
\includegraphics[scale=0.4]{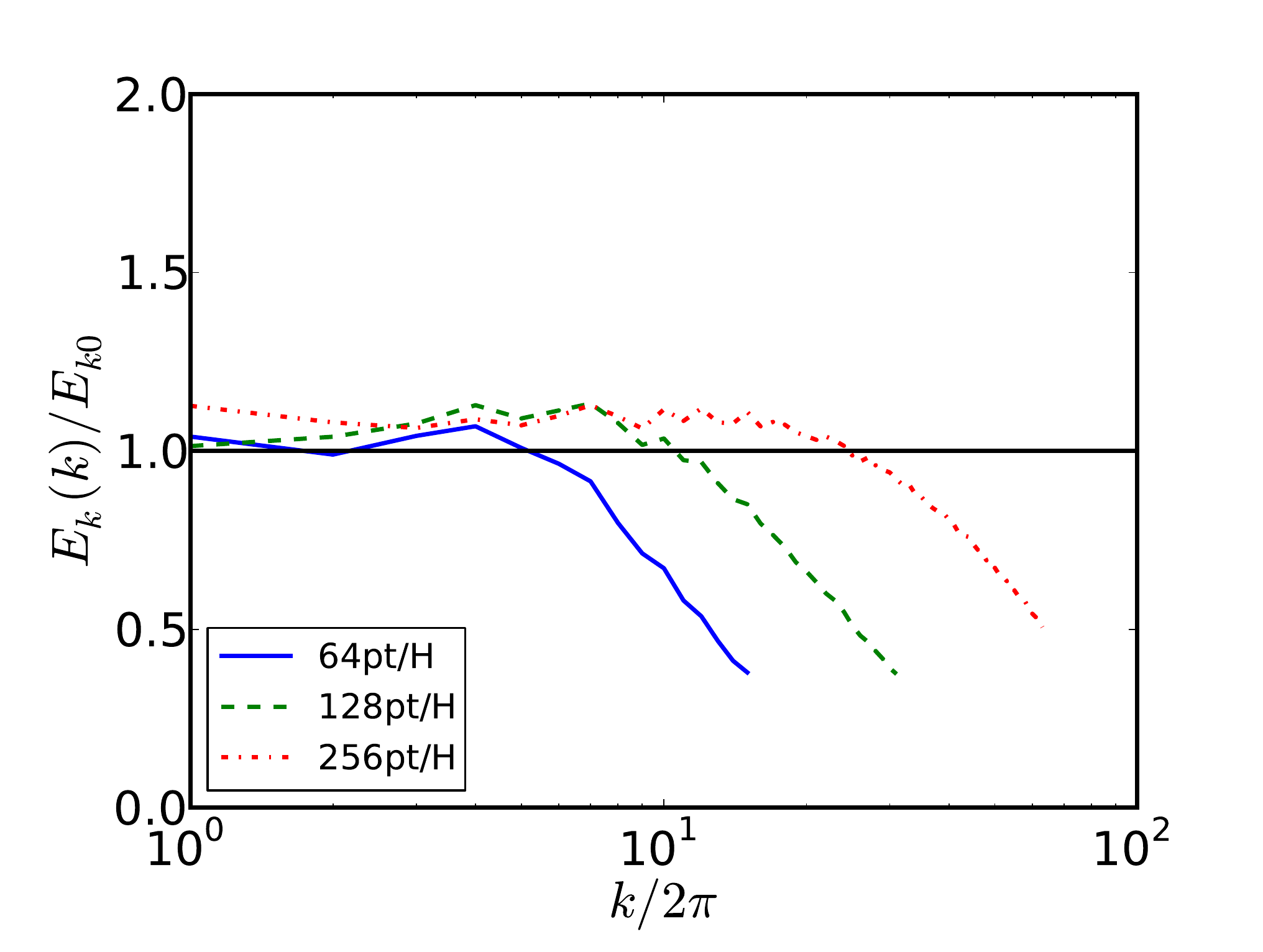}
\includegraphics[scale=0.4]{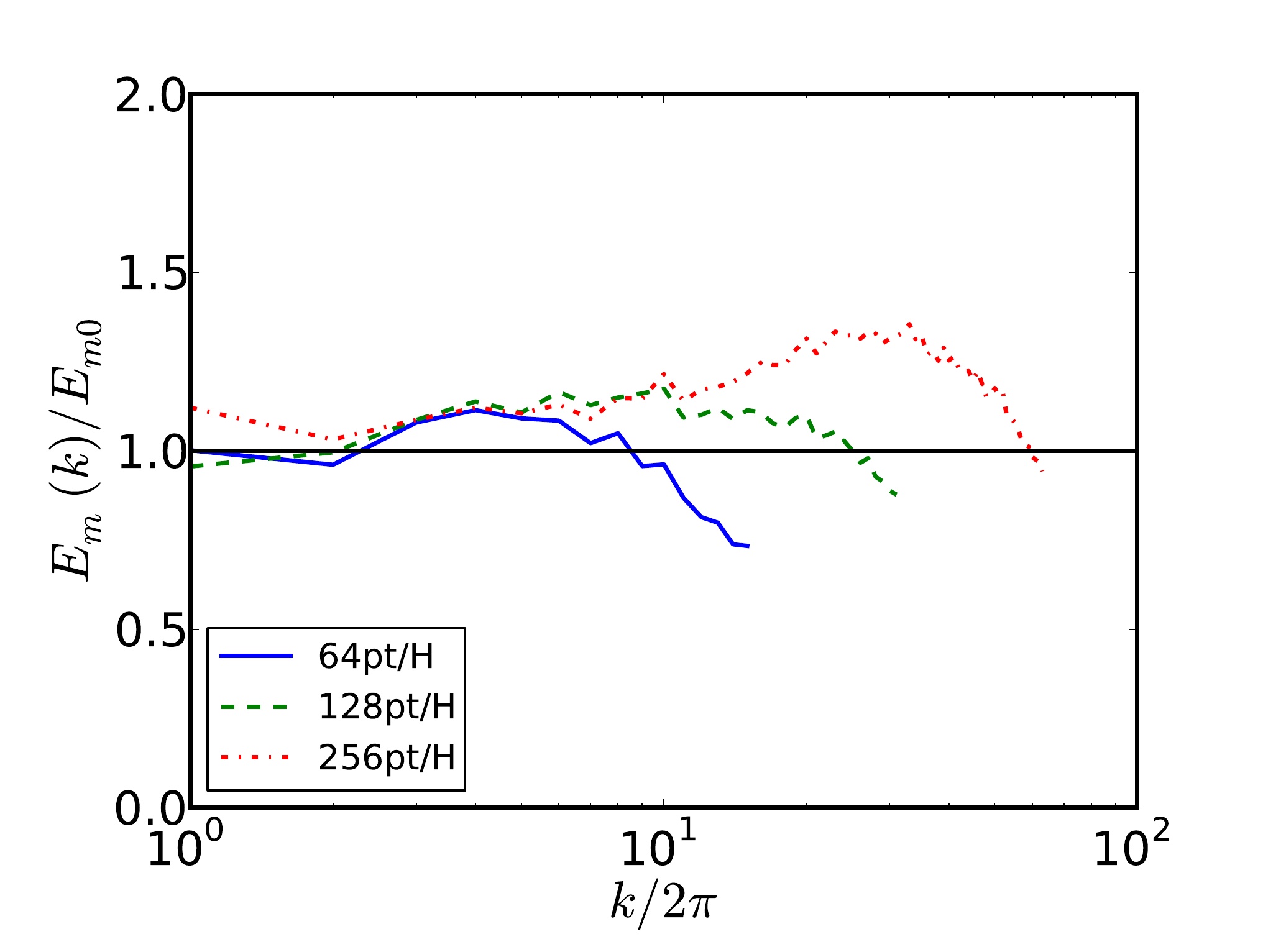}
\caption{Ratio between the power spectra obtained with
RAMSES in the ILES {\it (green, red, blue and dotted
curves)} and model Y-C-Re85000 in the azimuthal magnetic field
cases. Top panel is for kinetic energy power spectrum and 
bottom panel for magnetic energy power spectrum. On both panels, the
black horizontal line mark the location of unity ratio.
} 
\label{fig:Yles}
\end{center}
\end{figure}

\begin{figure}
\begin{center}
\includegraphics[scale=0.4]{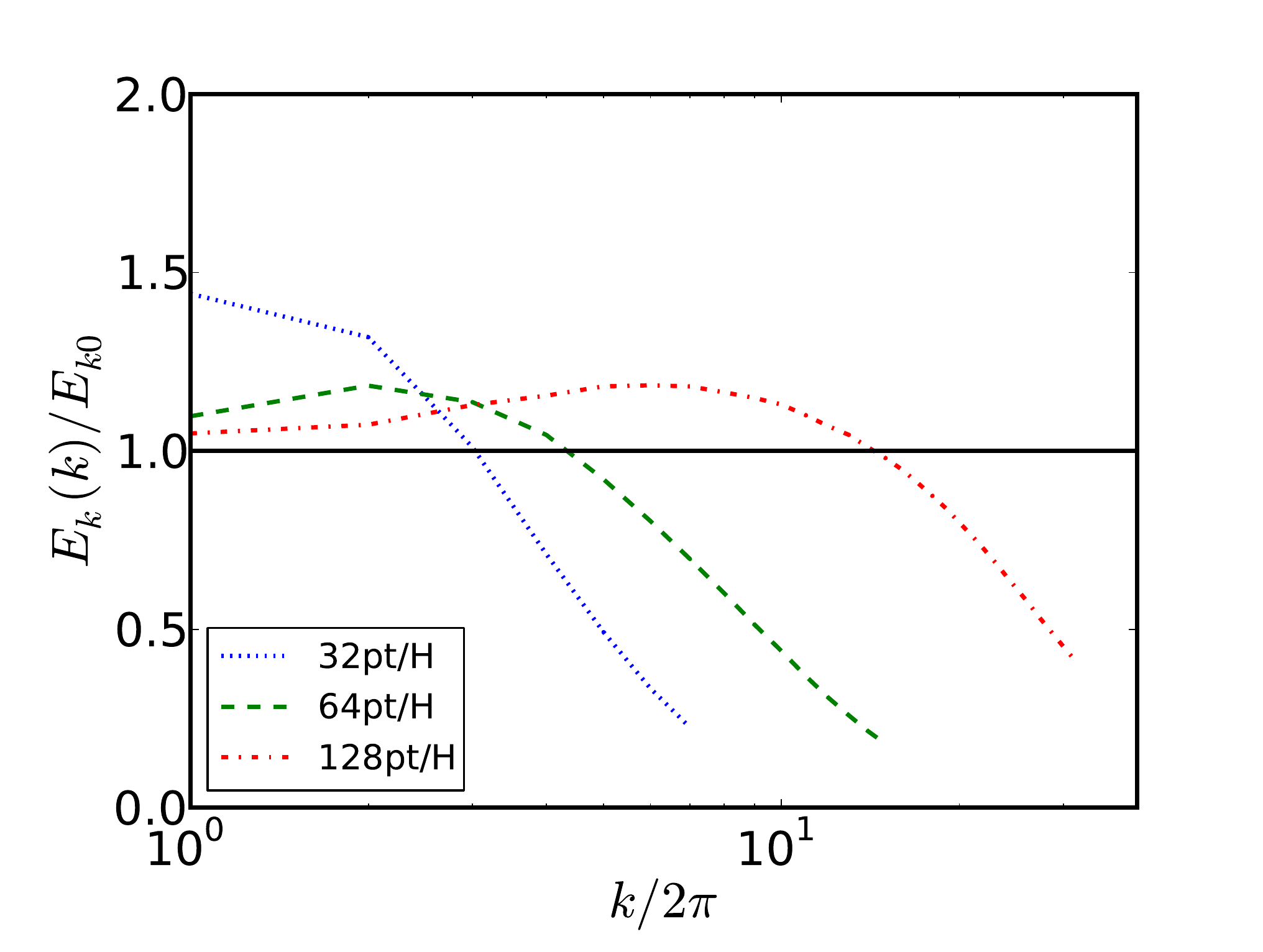}
\includegraphics[scale=0.4]{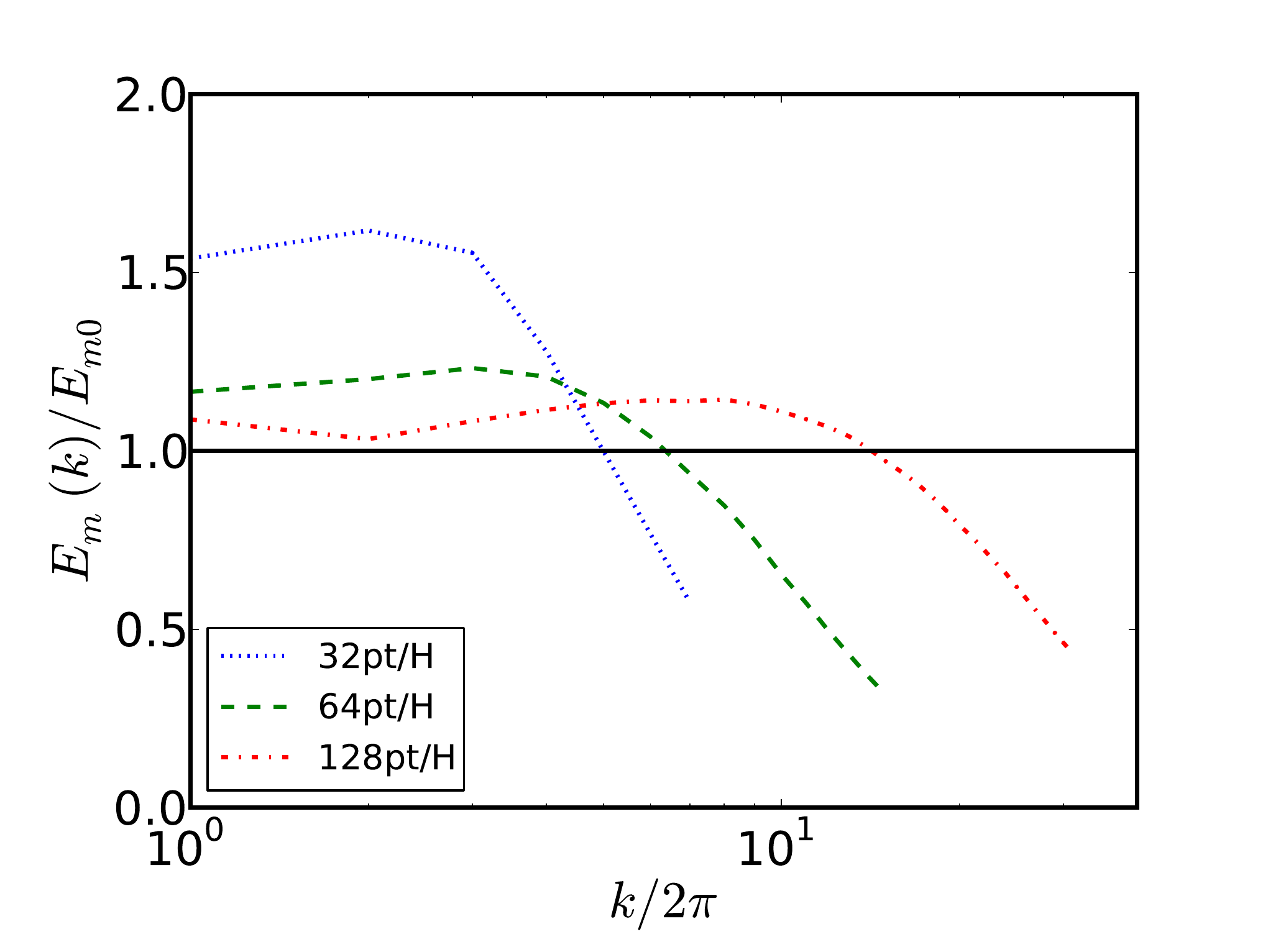}
\caption{Same as figure~\ref{fig:Yles}, but for the vertical field
  case.} 
\label{fig:Zles}
\end{center}
\end{figure}

In the small $Pm$ limit, the simplest possible sub--grid scale model
when using finite volume codes is certainly the
so--called ``Implicit LES'' (ILES). The idea is to capture Ohmic resistivity
explicitly in the simulation (since it occurs at large scale) but let
numerical dissipation handle kinetic energy dissipation at small
scales. This method has already been used in some previous works 
\citep{FSH00,IS05,OH11,FHK12,FRD15}, but its range of validity has never
been systematically and quantitatively investigated.
In this section, we compare the results of such simulations,
performed with RAMSES at various resolution with the results presented in
section~\ref{sec:DNS}. 

\paragraph{The azimuthal field case:} the magnetic
Reynolds number is here fixed to $Rm=2600$ as for the DNS. We performed a
series of simulations with resolutions ranging from $64$ to $256$
cells per scale height. For
all cases, we obtain $\alpha=1.8 \times 10^{-2}$, in very good
agreement with our best resolved simulations at $Pm=0.03$ (see
figure \ref{fig:alphaLES}). We then compared the kinetic and magnetic
energy power spectra in these simulations with the spectra obtained in
the DNS with the lowest Prandlt number (Y-C-Re85000). 
In the upper panel (resp. lower panel) of figure~\ref{fig:Yles},
we plot the ratio of the kinetic (resp. magnetic) power spectra
between the ILES and the DNS. 
The runs give an acceptable agreement with the DNS at all
available scales (except at small scales in the kinetic energy, which
is a result of the different hydrodynamical dissipation): 
at all resolutions, the deviations to the DNS run for $k'<10$ for
both the kinetic and the magnetic energy spectra, are at most of the order of $15 \%$.

\paragraph{The vertical field case:} the resolution in this case was
varied from $32$ cells per scale height to $256$ cells per scale
height. As can be seen on figure \ref{fig:alphaLES}, $\alpha$ decreases as
resolution increases, most likely as a result of the decrease of the effective
numerical viscosity, and converges to the value determined by the DNS simulations. Quite surprisingly though, the convergence as
resolution is increased toward
the asymptotic value of $\alpha$ at low $Pm$ is slower than the
azimuthal field case, despite $Rm$ being much larger in that
case. Indeed, with $64$ cells per scale height, the difference between
the two $\alpha$ values is still of order $20 \%$ while it has reached
convergence in the case of a mean $B_y$. This is probably
due to the stress dependence with $Pm$ being steeper in the presence
of a mean vertical field. In any case, $128$ cells per scale height
are needed to reach the asymptotic $\alpha$ value to less than $10
\%$. We note that this is still a gain of a factor of $16$ in
computing time. 
The power spectra
displayed in figure~\ref{fig:Zles} 
confirm that result: there is a difference of about $50 \%$ between
the power spectra of the ILES and the DNS at large scale for a
resolution of $32$ cells per scale height. Even the largest resolution
reveals differences of up to $20 \%$ in both the kinetic and magnetic
power spectra, even if the largest scales are converged to better
than $10 \%$, as expected from the good agreement between the
transport coefficients in both simulations (figure \ref{fig:alphaLES}).

%%%%%%%%%%%%%%%%%%%%%%%%%%%%%
%%%%%%%%%%Conclusions
\section{Conclusions}
\label{sec:conclusion}

We briefly summarise the main results and ideas of the paper and discuss some of their limitations. 

Our main goal was to investigate the properties of the turbulent state in the small magnetic Prandtl number limit by mean of local DNS simulations. We showed that in the presence of a mean magnetic field threading the domain, angular momentum transport converges to a finite value in the small $Pm$ limit. This result is valid both with a vertical and with an azimuthal mean magnetic field with the asymptotic values at small $Pm$ being respectively: $\alpha = 1.8\times 10^{-2}$ with $Rm=2600$ and $\beta_y\sim10^2$ and $\alpha = 3.2\times10^{-2}$ with $Rm=400$ and $\beta_z=10^3$.
The obtained values with our set of parameters are consistent with the estimations computed from the lifetime and accretion rate of protoplanetary discs ($\alpha$ is a few $10^{-2}$). 
Obviously, a word of care is in order here: the magnetization of accretion discs, such as protoplanetary discs, is only loosely constrained and the value of the $\beta$ parameter is unknown. 
$\alpha$ is known to strongly depend on the field strength, with $\alpha$ proportional to $\beta^{-1/2}$ \citep{HGB95} or $\beta^{-1}$ \citep{BCF11} in the vertical field case. Similarly, we can expect angular momentum transport to increase with the magnetic Reynolds number \citep{LL10}. The asymptotic values of the angular momentum rate we obtained are thus only valid for the $\beta$ and $Rm$ parameters we considered, and with a scaling that remains to be determined in the small $Pm$ limit. 

In the case of the compressible simulations with a mean vertical field, we obtained a surprising ratio of Maxwell to Reynolds stress of about $1$. Not withstanding any possible dependance with $\beta$ and $Rm$, we noticed that this ratio is related to the compressibility of the fluid, as a more usual value of about $3$ is obtained in the incompressible runs. It is also related to the bursty behavior of these two stresses: whereas a usual value is obtained in the growing phase of the bursts, the Reynolds stress dominates during its decreasing parts, resulting in a mean $R$ value of about $1$.  This result is specific to the compressible simulations, but it implies that neither the Reynolds stress nor the Maxwell stress converge to the same value in compressible and incompressible simulations at low $Pm$. In fact, both stresses differ by about 50\%, which we show can be attributed to their different behavior during the bursts. It is possible that the convergence of $\alpha$ at low $Pm$ to the same value with this two types of flow is fortuitous. In order to solve that issue, higher $Rm$ simulations are needed but are currently very costly.

Next, we investigated  the interest of using large eddy simulations both in spectral and real spaces for such simulations. The simplest approach is to consider the implicit LES method. As expected, in all cases the smallest scales are not correctly handled, but it is possible to reproduce the large scales energy spectra by using a high enough resolution. To obtain an error smaller than $20\%$ on $\alpha$, the needed resolution corresponds to a fourth of the resolution needed when explicit viscosity is included. This corresponds to a decrease in CPU time by a factor $256$. 
To limit the computational cost of a simulation of a turbulent flow, explicit SGS models are also usually considered.
The anisotropy of the rotating sheared flow and the turbulent cascade which differs from the Kolmogorov cascade, may indicate that simple SGS approaches are not well suited for MRI turbulence simulations.
We tested the Chollet-Lesieur method which is used in spectral space and is applied directly on the spectrum components. This method is local in frequency space, the effective viscosity being a function of the wave-number, and there is a limited inclusion of the backscatter. 
We found that good results are also obtained with this Chollet-Lesieur approach notwithstanding the anisotropy of the flow at small scales. 

Despite such positive results, a word of care is in order here. Although we showed that these methods are efficient to decrease the computational cost of MRI turbulence simulations, the needed resolution is still significant. For a magnetic Reynolds number $Rm=400$ with a vertical mean magnetic field, the required resolution is $64pts/H$. This $Rm$ is low compared to the values that are often chosen in the literature for which an even higher resolution is then necessary. To reach a turbulent flow dominated by Ohmic dissipation, a resolution of at least $256pts/H$ will be needed for $Rm$ values of a few thousands. Moreover, we considered only two diagnostics, namely the angular momentum transport rate and the power spectra, to reach this conclusion and other diagnostics, such as helicity or correlation functions, were not considered. Such statistical quantities may well be incorrectly described in our LES for the resolutions we used. More generally, with such reduced resolutions, any small scale process is not correctly handled, and for instance the study of collision of grains in protoplanetary discs or magnetic reconnection cannot be studied in such simulations.

Overall we still conclude that LES can be used to limit the computational time
of future simulations. Future works should account for density stratification, which will be particularly relevant in the context of global simulations. Large scale phenomena such as the ones identified in the `butterflies diagrams' can strongly modify the flow properties and are susceptible to affect the LES approach. Dedicated simulations should be performed to quantify the interest of such methods in this case.

\section*{ACKNOWLEDGMENTS}
HM, SF and MJ acknowledge funding from the
European Research Council under the European Union's Seventh Framework
Programme (FP7/2007-2013) / ERC Grant agreement n 258729. GL acknowledges
support by the European Community via contract
PCIG09-GA-2011-294110. The simulations presented in this paper were
granted access to the HPC resources of CCRT under the allocation
x2012042231 made by GENCI (Grand Equipement National de Calcul
Intensif).

\bibliographystyle{aa}
\bibliography{/Users/hmeheut/Documents/Recherche/Papers/biblioUTF8}

\end{document}